\AtBeginDocument{%
  \paperwidth=\dimexpr
    1in + \oddsidemargin
    + \textwidth
    + 1in + \oddsidemargin
  \relax
  \paperheight=\dimexpr
    1in + \topmargin
    + \headheight + \headsep
    + \textheight
    + 1in + \topmargin
  \relax
  \usepackage[pass]{geometry}\relax
}
\RequirePackage{fix-cm}
\documentclass[smallextended]{svjour3}  
\usepackage{adjustbox}
\usepackage{graphicx}
\usepackage{subcaption}
\usepackage{tcolorbox}
\usepackage{multirow}
\usepackage{comment}
\usepackage{tikz}
\usepackage{svg}
\usepackage{amsmath,amssymb,amsfonts}
\usepackage{pifont}
\usepackage{indentfirst}
\usepackage{url}
\usetikzlibrary{shapes.geometric, arrows}
\usepackage{algorithmic}
\usepackage{textcomp}
\usepackage{xcolor}
\usepackage{float}
\usepackage{tabularx}
\usepackage{listings}
\usepackage{framed}
\colorlet{shadecolor}{gray!10}
\colorlet{framecolor}{black}
\usepackage[numbers]{natbib}
\usepackage[inline]{enumitem}
\usepackage{makecell}
\usepackage{balance}
\usepackage{hhline}
\usepackage{lscape}
\usepackage{rotating}
\usepackage{soul}
\sethlcolor{green}
\usepackage[noframe]{showframe}
\usepackage{lipsum}
\definecolor{lightgray}{gray}{0.75}

\usepackage{hyperref}
\newcommand*\circled[1]{\tikz[baseline=(char.base)]{
            \node[shape=circle,draw,inner sep=1.5pt, scale = 0.9, fill=white, text=black] (char) {#1};}}
\newtcolorbox{rqbox}[1][]{
    colback=gray!10,
    colframe=gray,
    arc=1mm,
    boxrule=0.5pt,
    coltitle=black,
    fonttitle=\bfseries,
    title=#1
}
\begin{document}

\title{Towards Understanding the Impact of Data Bugs on Deep Learning Models in Software Engineering
}

\author{Mehil B. Shah         \and
         Mohammad Masudur Rahman \and 
         Foutse Khomh
}

\institute{Mehil Shah \at
              Dalhousie University, Canada \\
              \email{shahmehil@dal.ca}
           \and
           Mohammad Masudur Rahman \at
           Dalhousie University, Canada \\
           \email{masud.rahman@dal.ca}
           \and
           Foutse Khomh \at
           Polytechnique Montreal, Canada \\
           \email{foutse.khomh@polymtl.ca}
}

\date{Received: date / Accepted: date}

\maketitle

\begin{abstract}
Deep learning (DL) techniques have achieved significant success in various software engineering tasks (e.g., code completion by Copilot). However, DL systems are prone to bugs from many sources, including training data. Existing literature suggests that bugs in training data are highly prevalent, but little research has focused on understanding their impacts on the models used in software engineering tasks. In this paper, we address this research gap through a comprehensive empirical investigation focused on three types of data prevalent in software engineering tasks: code-based, text-based, and metric-based. Using state-of-the-art baselines, we compare the models trained on clean datasets with those trained on datasets with quality issues and without proper preprocessing. By analysing the gradients, weights, and biases from neural networks under training, we identify the symptoms of data quality and preprocessing issues. Our analysis reveals that quality issues in code data cause biased learning and gradient instability, whereas problems in text data lead to overfitting and poor generalisation of models. On the other hand, quality issues in metric data result in exploding gradients and model overfitting, and inadequate preprocessing exacerbates these effects across all three data types. Finally, we demonstrate the validity and generalizability of our findings using six new datasets. Our research provides a better understanding of the impact and symptoms of data bugs in software engineering datasets. Practitioners and researchers can leverage these findings to develop better monitoring systems and data-cleaning methods to help detect and resolve data bugs in deep learning systems.
\end{abstract}

\section {Introduction}

\looseness=-1
Deep Learning (DL) solutions have been widely adopted in many applications, including speech recognition~\cite{amodei2016deep}, software testing~\cite{mao2019extensive, cummins2018compiler}, autonomous driving~\cite{grigorescu2020survey}, and software development~\cite{yang2022dlinse}. However, deep learning models face unique challenges due to their complex architectures and data-driven paradigms~\cite{arpteg2018software}. Compared to traditional logic-driven software, DL-based software follows a data-driven computing paradigm where its models are trained using data. The training data can be noisy, biased, or incomplete, which may lead to unexpected or erroneous behaviours in the DL models~\cite{whang2020data}. Besides, deep learning models can be very complex and opaque, making it difficult to understand their decision process. Thus, the poor quality of training data in the DL models could pose significant challenges to their reliability and trustworthiness.

\looseness=-1
Debugging deep learning systems is a challenging task. DL bugs can originate from various sources, such as training data, hyperparameters, and model structure, and can lead to system crashes and unexpected runtime behaviour~\cite{islam2019fse}. They can also result in severe consequences, as evidenced by the fatal accidents involving self-driving cars~\cite{selfdrivingcarcrash, teslaAutopilotCrash}. The non-determinism of DL systems leads to different outcomes across multiple runs, making debugging challenging~\cite{nagarajan2018impact}. According to existing literature~\cite{islam2019fse}, DL bugs can be divided into five categories: Data, Model, Structural, Non-Model Structural and API Bug.

\looseness=-1
Among these categories, data bugs have been reported as the most prevalent ones, accounting for 26\% of all bugs in deep learning systems~\cite{islam2019fse}. They originate from errors in training data, such as incorrect labels, duplicates, out-of-distribution records, and missing values~\cite{islam2019fse, taxonomyRealFaults, cote2024data}. These errors can significantly affect the model performance~\cite{liang2022advances, sambasivan2021everyone}. Resolving data bugs is very challenging~\cite{yin2023dynamic, wang2024empirical}. Data bugs are hidden in the dataset and implicitly affect a model's behaviour. They also propagate to the model parameters during training, which makes their detection difficult. Furthermore, according to existing work~\cite{croft2022data}, many existing benchmark datasets constructed by human annotators contain up to 70\% mislabeled data, which leads to data bugs in the DL models relying on those datasets~\cite{croft2022data}.

\looseness=-1
Existing studies~\cite{tantithamthavorn2015impact, mahmood2015impact, kim2011dealing, wu2021data, xu2023data, croft2023data} have investigated the impact of data quality issues (e.g., label noise, class imbalance) on model performance. However, they primarily focus on the quality of the raw training data and overlook two crucial aspects: the preprocessing stage and the model training process. First, data preprocessing is essential for preparing the data in a suitable format for training. Errors, biases, or loss of information during preprocessing can propagate through the whole development steps of a model, significantly degrading the model's performance~\cite{amato2023data}. Despite its importance, the impact of preprocessing on model performance has not been thoroughly investigated by the existing literature. Second, training is a crucial step in model development. Observing the training process and monitoring a model's internal state, such as gradients, weights, and biases, can offer valuable insights into how data quality issues affect the model's learning process. For example, if the model is experiencing vanishing or exploding gradients, they might indicate poor feature scaling or outliers in the data~\cite{glorot2010understanding}. Similarly, analyzing the weights and biases of the model during training can indicate whether the model is learning meaningful patterns or is being misled by noisy or corrupted data points~\cite{koh2017understanding}. Such insights can help practitioners take appropriate actions to mitigate the impact of data bugs in their DL-based applications. However, no existing work examines the impact of data bugs on the training behaviours of DL models. Our study aims to address this critical gap in the literature.

\looseness=-1
In this paper, we conduct an empirical study to investigate the impact of data quality and preprocessing issues on the training of deep learning models used in software engineering tasks. First, we select three types of data for our analysis based on their frequent use in software engineering tasks: code-based, text-based, and metric-based. Second, we select state-of-the-art baseline models using these data types and compare their faulty models (containing data bugs) with corresponding bug-free versions. Third, we capture the detailed training logs using the Weights \& Biases framework and analyze the training metrics and model properties (e.g., gradients, weights, and biases) to derive meaningful insights. We perform qualitative and quantitative analyses of the gradients, weights, and biases to identify symptoms and manifestations of data quality and preprocessing issues. Finally, we validate our findings with new datasets to ensure the generalizability of our results. Thus, we answer four important research questions as follows:

\textbf{RQ1}: \textit{How do data quality and preprocessing issues in code-based data affect the training behavior of deep learning models?} 

We investigate the impact of data quality issues from two code-based datasets, Devign and BigVul~\cite{zhou2019devign, fan2020ac}. These datasets contain C/C++ functions from multiple projects and have known quality issues~\cite{croft2023data}. Our analysis reveals that these issues cause gradient instability during training and significantly affect the weight and bias distribution of the deep learning models. Furthermore, our manual analysis of attention weights shows that data quality issues in the code-based data can reduce the code comprehension abilities of the deep learning model.

\textbf{RQ2}: \textit{How do data quality and preprocessing issues in text-based data affect the training behaviour of deep learning models?}

We assess the impact of data quality issues from two text-based datasets, Eclipse, maintained in Bugzilla and Hadoop, maintained in JIRA. Our study reveals that quality issues in text data lead to reduced abnormal weight distributions and overfitting of models to noisy patterns during model training. Our qualitative analysis using the t-SNE plots highlights the deep learning models' difficulty learning consistent feature representations from noisy data. 

\textbf{RQ3}: \textit{How do data quality and preprocessing issues in metric-based data affect the training behavior of deep learning models?}

We investigate the impact of data quality issues from two metric-based datasets, Openstack and QT, which frequently suffer from class imbalance problems~\cite{mcintosh2018fix, hoang2019deepjit}. Our analysis reveals that quality issues in metric-based data lead to vanishing gradients and higher loss during the training process. Additionally, we perform qualitative analysis using GradCAM, an explainable AI technique for visual analysis of model inputs. Our qualitative analysis demonstrates that models trained on imbalanced data focus on irrelevant tokens and struggle to generalize to unseen data.

\textbf{RQ4}: \textit{How well do our findings on data quality and preprocessing issues generalize to other code-based, text-based, and metric-based datasets?}

\looseness=-1
We evaluate the generalizability of our findings using six new datasets: D2A and Juliet (code-based), Spark and Mozilla (text-based), and Go and JDT (metric-based). Our analysis reveals that the data quality and processing issues in the training data lead to near-zero biases, smaller weights, gradient instability, and overfitting, which align with our above findings. In contrast, the models trained on cleaned datasets show none of these issues. Such observations increase confidence in our findings and underscore the challenges of data bugs in deep learning models used in software engineering tasks.

\underline{\textbf{Paper Organization:}} The remainder of the paper is organized as follows. Section 2 provides background knowledge about deep learning and explainable AI. Section 3 describes the methodology of our study in detail, including data type selection, data collection, experimental setup, post-hoc analysis and quantitative analysis. Section 4 talks about the findings for the different research questions. Section 5 discusses the implications of our findings for various stakeholders. Section 6 reviews the related literature. Finally, Section 7 presents threats to the validity of our study, and Section 8 concludes the paper.

\section {Background}

\subsection{Data Bug}
A \textbf{data bug} is a systematic error or flaw in the data pipeline that leads to incorrect, inconsistent, or misleading data, thereby compromising the integrity or performance of a machine learning or data-driven system~\cite{islam2019fse}. Formally, for a data pipeline \( \mathcal{P} \) that transforms raw data \( X_{\text{raw}} \) into processed data \( X_{\text{proc}} \), a data bug exists if:
\[
\exists\ \mathcal{E} \in \mathcal{P}\ \text{such that}\ \mathcal{E}(X_{\text{raw}}) \neq X_{\text{expected}},
\]
where \( X_{\text{expected}} \) is the correct output of the pipeline under ideal conditions. Data bugs can arise from issues such as data collection from unreliable sources, faulty transformations, incorrect feature engineering, erroneous dimensionality reduction, and inappropriate data type conversions that alter the fundamental nature of the data. They may also arise from missing preprocessing steps, such as failure to handle missing values, overlooking duplicate entries, and neglecting the inconsistencies in the data.

\textbf{Class Imbalance}:
Class imbalance refers to a scenario where the distribution of classes in the dataset is highly skewed. Formally, given a dataset \( \mathcal{D} = \{(x_i, y_i)\}_{i=1}^N \) with \( y_i \in \{1, 2, ..., K\} \), class imbalance exists if the class prior probabilities \( P(Y=k) \) satisfy:
\[
\exists\ k, l \in \{1, ..., K\}\ \text{such that}\ \frac{P(Y=k)}{P(Y=l)} \geq \tau,
\]
where \( \tau \gg 1 \) is a threshold indicating significant disparity (e.g., \( \tau = 10 \) for severe imbalance). For instance, in software engineering domains, moderate imbalance is characterized by \( \tau \approx 1.5 \) (corresponding to ratios like 60:40) and might be acceptable. However, severe imbalance occurs at \( \tau \geq 3 \) (corresponding to ratios of 75:25 or greater), which is undesirable~\cite{mahbub2023defectors,li2022robust}.

\textbf{Concept Drift}:
Concept drift refers to a change in the underlying data distribution over time, specifically in the relationship between input features \( X \) and the target variable \( Y \). For a time-varying distribution \( P_t(X, Y) \), concept drift occurs between times \( t_1 \) and \( t_2 \) if:
\[
P_{t_1}(Y \mid X) \neq P_{t_2}(Y \mid X),
\]
even if \( P_{t_1}(X) = P_{t_2}(X) \). This distinguishes it from \textit{covariate shift} (where only \( P(X) \) changes).

\textbf{Label Noise}:
Label noise arises when training labels are corrupted. For a true labeling function \( f: \mathcal{X} \rightarrow \mathcal{Y} \), observed labels \( \tilde{y}_i \) are perturbed such that:
\[
\tilde{y}_i = 
\begin{cases} 
y_i & \text{with probability } 1-\eta, \\
z \sim \mathcal{Y} \setminus \{y_i\} & \text{with probability } \eta,
\end{cases}
\]
where \( \eta \in [0, 1] \) is the noise rate~\cite{frenay2014comprehensive}. Label noise can be \textit{uniform} (random flips) or \textit{structured} (biased toward specific classes).

\textbf{Out of Distribution Data}: 
Out-of-distribution (OOD) data refers to data that does not come from the probability distribution of the training data. A trained model is assumed to generalize well to new data from the training distribution. OOD data violates this assumption and could lead to poor model performance.

Formally, given a training distribution $P_{\text{in}}(X,Y)$ and a test distribution $P_{\text{out}}(X,Y)$, a data point $(x,y)$ is considered out-of-distribution if it is drawn from $P_{\text{out}}$ and $P_{\text{out}} \neq P_{\text{in}}$.

The core challenge with OOD data is that a model, having learned patterns specific to the training distribution, may produce overconfident or incorrect predictions for OOD samples, as it has not been exposed to similar examples during training. While OOD characteristics can arise from a natural domain shift, they become a data quality issue when a defect or an error causes a significant deviation from the training distribution. For example, consider a model trained to classify images of cars. If the model is trained only on images of a specific type of car (e.g., sedans), a new set of images of another type of car (e.g., pickup trucks), can be considered as out-of-distribution data due to a natural distributional shift. However, if the new images contain corrupt data (e.g., solid black images) or incorrect labels, this out-of-distribution deviation represents a data quality issue caused by errors in preprocessing or manual annotation.

\textbf{Missing Preprocessing Operations}:
Missing preprocessing operations occur when input data deviates from the expected distribution assumed by a deep learning (DL) model due to omitted transformations. Let \( T: \mathcal{X} \rightarrow \mathcal{X}' \) denote a required preprocessing function (e.g., normalization, imputation). If raw data \( X \) is fed to a model trained on \( T(X) \), the input distribution mismatch degrades the model's performance. Formally, the model \( M \) optimized for \( P(T(X)) \) faces a distribution shift when applied to \( P(X) \), violating the i.i.d. assumption. In other words, the missing preprocessing operations introduce a shift in the distribution between the data the model was trained on and the data it encounters during inference. This shift in the distribution violates the assumption that the distributions during training and inference will be identical, and this causes a drop in the model's performance.

\subsection{Data Quality Issues in Software Engineering Datasets}

\textbf{Label Noise:} Label noise refers to errors in the labels assigned to data instances in a dataset. These errors can stem from various sources, including insufficient information, annotator mistakes, subjective judgments, and data encoding issues~\cite{frenay2014comprehensive}. Label noise is prevalent in real-world datasets and thus can significantly impair the DL models using those datasets~\cite{frenay2013classification}. This issue is also relevant in software engineering datasets since they are often used to train DL models supporting code search, vulnerability detection, and program understanding~\cite{wang2024empirical, xu2024code, nie2023understanding}. For example, in a dataset for vulnerability detection, non-vulnerable code could be mistakenly labelled as vulnerable or vulnerable code could be labelled as non-vulnerable. Such label errors often result in inaccurate, biased, or flawed models~\cite{nie2023understanding}. In this work,  we train multiple deep learning models using datasets containing label noise to determine its impact on the models.

\textbf{Class Imbalance:} Class imbalance refers to a disproportionate representation of different classes within a dataset. It is also highly prevalent in various software engineering datasets, including API recommendation, code review automation, and defect prediction~\cite{irsan2023multi, tufano2022using, song2018comprehensive}. For example, in defect prediction, there are often significantly fewer defective modules than non-defective ones. Existing studies~\cite{mcintosh2018fix, kamei2012large, keshavarz2022apachejit, mahbub2023defectors} show that only 5\%-26\% of the files contain the defective instances. When trained on imbalanced datasets, DL models tend to be biased towards the majority class and demonstrate poor performance in identifying the minority class~\cite{giray2023use}.

\looseness=-1
\textbf{Data Obsolescence:} Data obsolescence, also known as concept drift, refers to the evolution of data over time and is a significant challenge for software engineering datasets. As software systems evolve, the characteristics of their data change, which leads to concept drift. This phenomenon is prevalent in many datasets, including the ones used for log-level prediction, anomaly detection and duplicate bug report detection~\cite{ma2018robust, ouatiti2024impact, zhang2023duplicate}. Recent research by Zhang et al.~\cite{zhang2023duplicate} revealed that most duplicate bug report detection techniques were only evaluated using data up to January 2014. When these same techniques were applied to more recent data, their performance decreased significantly, highlighting the impact of data obsolescence on the effectiveness of deep learning models.

\subsection{Preprocessing Faults in Deep Learning Systems}

In this section, we discuss the most common types of preprocessing faults in deep learning systems, as per the existing taxonomy by Humbatova et al.~\cite{taxonomyRealFaults}. These faults broadly manifest in two categories: missing preprocessing and incorrect preprocessing.

\textbf{Missing Preprocessing:} Missing preprocessing refers to a scenario when a preprocessing step leading to better performance has not been applied at all. As suggested by the existing literature~\cite{taxonomyRealFaults}, missing preprocessing can significantly impact the model performance. For code analysis, cases were reported where the lexer and parser failed to catch certain code structures due to missing preprocessing steps. In text processing, particularly with social media data and software engineering text, issues arose from not handling special characters and different encodings (UTF-8, Unicode). With metric-based data, the absence of correlation analysis between features led to redundant features in the training data, and missing normalization for numerical values affected model performance.

\textbf{Incorrect Preprocessing:} Incorrect preprocessing refers to a scenario when a preprocessing step was either inappropriate or incorrect. According to existing literature~\cite{taxonomyRealFaults}, incorrect preprocessing can manifest in subtle ways. In code preprocessing, incorrect token replacement and lexer implementations led to improper parsing of programming constructs. For text data, developers encountered problems with incorrect handling of special tokens, where preprocessing tokens were incompatible with the existing toolchain. One specific case involved the Stanford Parser producing incorrect tags for certain words, leading to the wrong interpretation of the data. For metric-based features, when different features of the numerical data were scaled inconsistently across different ranges (e.g., [0,1] versus [-1,1]), this resulted in performance degradation~\cite{taxonomyRealFaults}.

In this study, we focus exclusively on missing preprocessing faults for several reasons. First, missing preprocessing faults occur more frequently in deep learning systems than incorrect preprocessing~\cite{taxonomyRealFaults}. Second, missing preprocessing can be systematically studied by removing the preprocessing steps. However, there currently exists no systematic approach for simulating wrong preprocessing, as existing mutation techniques focus primarily on training, model, and data faults~\cite{humbatova2021deepcrime}. Third, manually selecting the incorrect preprocessing operations could introduce bias, leading to unreliable findings.

\subsection{Model Explanation}

\textbf{Attention-Based Analysis}: Attention mechanisms help deep learning models focus on the most relevant parts of the input~\cite{vaswani2017attention}. Their inherent ability to assign importance weights to different input elements can also be leveraged to interpret the behaviours of the models. Recently, attention weights of input have been used to investigate the explainability of the deep learning models solving software engineering tasks~\cite{fu2022linevul, pornprasit2022deeplinedp, fu2022gpt2sp}. For example, Fu et al.~\cite{fu2022linevul} leveraged the self-attention mechanism to explain the predictions of their proposed technique for vulnerability detection. In this study, we employ their dataset and attention-based analysis to examine how data quality and preprocessing issues affect a DL model's training behaviour and learning capacity.

\textbf{t-SNE}: t-Distributed Stochastic Neighbor Embedding (t-SNE)~\cite{van2008visualizing} is a technique that can reduce high-dimensional feature representations learned by neural networks and visualize them in 2D or 3D plots. These plots are often used to inspect how well a model has learned to differentiate among different classes in the data. By comparing t-SNE visualizations for different models and examining their separation of classes, we can evaluate their capacities to learn feature representations for a given task. This technique has also been used in several software engineering tasks, including duplicate bug report detection~\cite{messaoud2022duplicate}. As duplicate bug report detection is one of our selected tasks, we utilise t-SNE to explain and visualize the predictions of our trained model for this task.

\looseness=-1
\textbf{Grad-CAM}: Grad-CAM (Class Activation Mapping) is a technique that visualizes the input features contributing the most to a neural network's outputs~\cite{selvaraju2017grad}. By analyzing the feature weights in the final convolutional layer of a convolutional neural network, Grad-CAM produces a heatmap highlighting the important regions in the input relevant to the model's output. In our study, we employ DeepJIT~\cite{hoang2019deepjit} as a baseline technique for defect prediction. Since DeepJIT is a CNN-based technique, we use Grad-CAM to understand and visualize the impact of input features on the model output.

\section {Methodology}

Fig.~\ref{fig:schematicDiagram} shows the schematic diagram of our empirical study. We discuss different steps of our study as follows.

\subsection{Data Type Selection}
Selecting different types of data is crucial for our study since each data type has unique characteristics and quality issues. By examining multiple types of data, we can comprehensively investigate how data bugs manifest and affect deep learning models. Based on the prevalence in software engineering datasets, Yang et al.~\cite{yang2022dlinse} identified the three most prevalent types of data: code-based, text-based, and metric-based. These data types have unique characteristics as follows.

\textbf{Code-based data:} Code-based data is frequently used in training deep learning models that target various software engineering tasks, such as code clone detection, code generation, program repair, and vulnerability detection~\cite{fu2022linevul, zhang2020learning, li2017cclearner, tian2020evaluating}. This data type encompasses source code files, test cases, and code changes~\cite{yang2022dlinse}. 

\textbf{Text-based data:} Natural language texts play a crucial role in numerous software engineering tasks. Existing studies have used text-based data in deep learning techniques for software engineering tasks~\cite{yang2022dlinse}. Text-based data includes requirements specifications, design documents, code comments, commit messages, bug reports, user reviews, and question-answer posts from forums like Stack Overflow~\cite{he2020duplicate, ouatiti2024impact, mondal2024can}.

\textbf{Metric-based data:} Metric-based data comprises various statistics derived by static analysis tools (e.g., SonarQube, Understand) from various software repositories. They quantify different aspects of source code, software design, development process, and software quality. Metric-based datasets have been used in several software engineering tasks such as defect prediction, effort estimation, code smell detection, and software maintainability assessment~\cite{hoang2019deepjit, liu2019deep, choetkiertikul2018deep}. 

Thus, based on the prevalence of data types and their relevance to software engineering tasks, we consider \textit{code-based}, \textit{text-based} and \textit{metric-based} data for our study (Step \circled{1}, Fig.~\ref{fig:schematicDiagram}). 

\subsection{Study Design}
\subsubsection{Task Selection}
\begin{figure*}
\centering
\includegraphics[width=\textwidth]{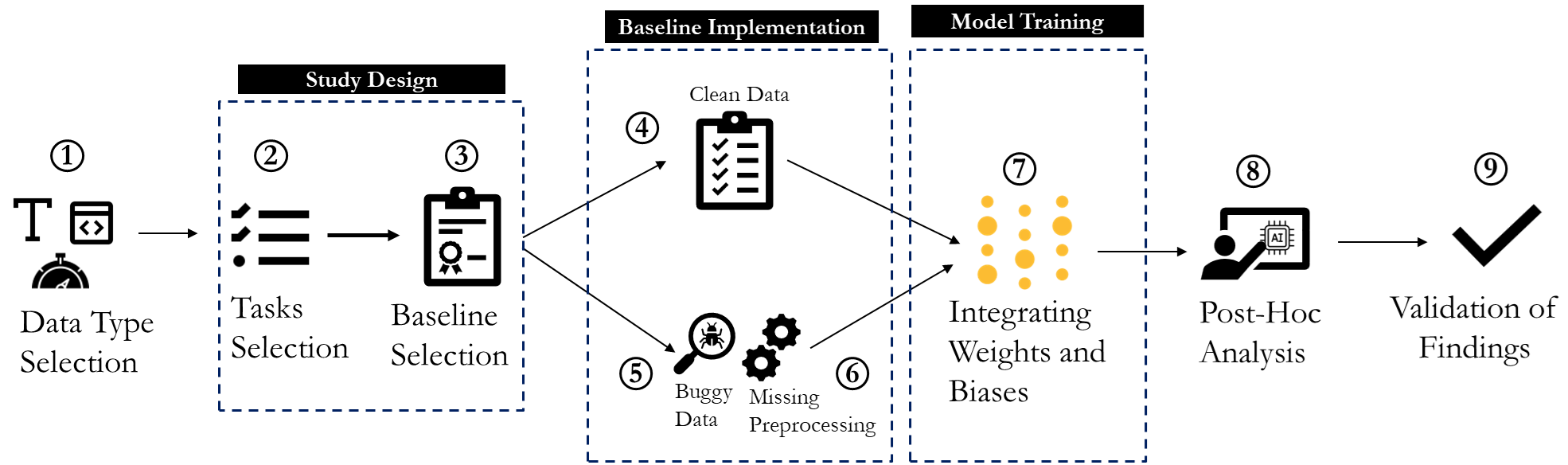}
\caption{Schematic diagram of our study}
\label{fig:schematicDiagram}
\end{figure*}

Selecting representative tasks for each type of data above is a critical step for our analysis (Step \circled{2}, Fig.~\ref{fig:schematicDiagram}). By carefully choosing the tasks that use our selected data types, we can examine how data bugs manifest and affect the training of deep learning models. In particular, we have selected classification tasks for our analysis due to their several benefits. First, classification-based tasks tackle many important challenges in software engineering (e.g., vulnerability detection, bug localization, duplicate bug report detection, code clone detection), leveraging different types of data. On the other hand, generation tasks in software engineering are primarily focused on code~\cite{yang2022dlinse}. Thus, selecting generation tasks could limit our ability to study the impact of data quality and preprocessing issues across different types of data. Second, classification tasks provide straightforward evaluation metrics (e.g., accuracy, precision, recall) for measuring the impact of data quality issues, which is crucial for our analysis. Thus, we have selected the following representative tasks for our analysis, as suggested by Yang et al.~\cite{yang2022dlinse}:

\textbf{(a) Vulnerability Detection:} Detection of vulnerabilities in software code is one of the key applications where deep learning models show promising results. These models often use source code or binary for their detection task.

\textbf{(b) Duplicate Bug Report Detection:} Duplicate bug report detection is a prominent use case of DL models leveraging text-based data from software engineering. The task's objective is to identify the incoming bug reports that describe the same underlying issue from the past bug reports.

\textbf{(c) Defect Prediction:} Predicting defects in code components (e.g., classes, methods) using software metrics is a common application of deep learning in software engineering. The task involves predicting the defect proneness of software modules based on various code metrics (e.g., complexity, coupling, cohesion). 

\subsubsection{Baseline Selection}
After selecting the representative tasks, we choose multiple baseline approaches for them. These approaches provide baseline deep learning models, which are used in our subsequent analysis. To select baselines for each task, we used the following criteria: (a) the technique should use a neural network model to perform one of the three tasks above, and (b) the technique should provide a comprehensive replication package with source code, datasets, and instructions for reproducibility. We also look for baselines with clean and buggy versions of the data in their replication packages. Based on these filtration criteria, we selected the following state-of-the-art baselines for our study.

\textbf{(a) Code-based data: LineVul~\cite{fu2022linevul}, CodeBERT~\cite{ding2024vulnerability}}: LineVul and CodeBERT are state-of-the-art transformer-based models for line-level vulnerability detection in code. The models take code tokens as input and learn to predict the vulnerability status of each line.

\textbf{(b) Text-based data: DCCNN~\cite{he2020duplicate}, BERT-MLP~\cite{messaoud2022dbrd}}: DCCNN, short for Duplicate bug report detection using Convolutional Neural Networks, employs a CNN-based architecture to identify duplicate bug reports leveraging their textual descriptions. The CNN model first learns to extract relevant features and patterns from a pair of bug reports and then predicts whether they are duplicates. BERT-MLP, a representation learning-based technique, uses the BERT model to learn the contextual relationship between words. Then, it uses a softmax-based classifier to determine if two bug reports are duplicates or not based on their representation.

\looseness=-1
\textbf{(c) Metric-based data: DeepJIT~\cite{hoang2019deepjit},  CodeBERTJIT~\cite{guo2023jit}}: DeepJIT, and CodeBERTJIT are end-to-end deep learning frameworks for Just-In-Time (JIT) defect prediction. These techniques compute software metrics from datasets of code changes and extract salient features from commit messages. DeepJIT uses CNN to extract the salient features, whereas CodeBERTJIT utilizes CodeBERT to extract the features. The learned features from commit messages and code changes are encoded into numerical matrices and then processed by separate CNN layers to predict whether the commit will likely introduce a defect. 

Besides strong relevance to our selected data types, these baseline techniques have demonstrated state-of-the-art performance in their respective tasks, which justifies their selection for our study (Step \circled{3}, Fig.~\ref{fig:schematicDiagram}).

\subsubsection{Datasets}
We utilized six diverse datasets that were used by deep learning models targeting our tasks above: Devign and BigVul (code-based), Eclipse and Hadoop (text-based), and OpenStack and QT (metric-based). These datasets vary significantly in size, composition, and specific characteristics, allowing for comprehensive training and evaluation of deep learning models across different tasks. For each type of data, we chose two datasets to perform data source triangulation, per the guidelines by Runeson et al.~\cite{runeson2009guidelines}, and to have multiple sources of evidence as recommended by Yin et al.~\cite{yin2009case}. Table.~\ref{tab:summaryDatasets} provides a detailed summary of each dataset, including their sizes and brief descriptions, offering a clear overview of the data used in our experiments.
\begin{table}[ht]
\centering
\caption{Summary of datasets used in our study}
\label{tab:summaryDatasets}
\resizebox{\linewidth}{!}{%
\begin{tabular}{|l|l|l|l|}
\hline
\textbf{Data Type} &
  \textbf{Dataset} &
  \textbf{Size} &
  \textbf{Description} \\ \hline
\multirow{2}{*}{Code-Based} &
  Devign~\cite{zhou2019devign} &
  27,318 functions &
  Vulnerable functions from 5 open-source projects in C/C++ \\ \cline{2-4} 
 &
  BigVul~\cite{fan2020ac} &
  188,636 functions &
  Vulnerable code snippets from 211 projects in multiple languages \\ \hline
\multirow{2}{*}{Text-Based} &
  Eclipse~\cite{lazar2014generating} &
  74,376 bug reports &
  Bug reports from Eclipse project (2001-2014) \\ \cline{2-4} 
 &
  Hadoop~\cite{hadoop} &
  14,016 bug reports &
  Bug reports from Hadoop project (2012-2014) \\ \hline
\multirow{2}{*}{Metric-Based} &
  OpenStack~\cite{mcintosh2018fix} &
  66,065 source files &
  21 code metrics per file from OpenStack project \\ \cline{2-4} 
 &
  QT~\cite{mcintosh2018fix} &
  95,758 source files &
  21 code metrics per file from QT project \\ \hline
\end{tabular}%
}
\end{table}

\subsubsection{Quality Issues in Existing Datasets}
The benchmark datasets utilized in our analysis exhibit several quality concerns that warrant discussion:

\begin{itemize}
\item \textbf{Vulnerability Detection (Devign, BigVul)}: Prior research has identified significant quality issues in these datasets~\cite{croft2023data}. Existing analysis indicates that only 54\% and 80\% of the samples in Big-Vul and Devign maintain correct labels, demonstrating substantial label noise~\cite{croft2023data}. Furthermore, these datasets face challenges with temporal relevance, as 15\% of Devign samples and 25\% of Big-Vul samples no longer reflect the current trends of vulnerability~\cite{croft2022data}.

\item \textbf{Duplicate Bug Report Detection (Hadoop, Eclipse)}: These datasets exhibit similar quality concerns regarding data representation and temporal validity. Studies have shown that only 5.2\% of Eclipse bug reports and 2.7\% of Hadoop bug reports are marked as duplicates~\cite{zhang2023duplicate}. Besides, the majority of their bug reports were submitted in 2014 or earlier, potentially leading to concept drift problems in these datasets~\cite{zhang2023duplicate}. Research has also demonstrated that approximately 25-40\% of bug reports in Bugzilla-based systems contain incorrect labels, indicating substantial label noise in these systems~\cite{tantithamthavorn2015impact}.

\item \textbf{Defect Prediction (OpenStack and QT)}: The OpenStack and QT datasets present multiple quality challenges. Research has revealed that their dataset construction methodology may introduce significant labelling errors, potentially affecting up to 55\% of the data~\cite{yatish2019icse}. These datasets also exhibit major class imbalance, with defective lines constituting only 20\% of the total samples~\cite{hoang2019deepjit}. Additionally, these datasets were constructed using commits before 2014, which might introduce significant concept drift in the techniques leveraging these datasets~\cite{hoang2019deepjit}.
\end{itemize}

\subsection{Experimental Setup}
This section describes our experimental methodology, including how we configure our system, implement the baselines, and observe training behaviours.

\subsubsection{System Configuration}
To reflect the original environments of our selected baselines, we use the following setup:

\textbf{(a) Code Editors:} We use Visual Studio Code v1.79.0, which is a popular code editor for building DL-based applications~\cite{ide}.

\textbf{(b) Dependencies:} To automatically detect the API libraries used in the baseline techniques, we use the pipreqs package\footnote{https://pypi.org/project/pipreqs/}. We also install the dependencies for each baseline into a separate virtual environment using the venv\footnote{https://docs.python.org/3/library/venv.html} module.

\looseness=-1
\textbf{(c) Frameworks:} We use Tensorflow and PyTorch for our experiments, as DeepJIT~\cite{hoang2019deepjit} and LineVul were originally developed in PyTorch, whereas DCCNN was originally developed in Tensorflow.

\textbf{(d) Python Version:} For our experiments, we use the same versions of Python originally used by the authors when they published their work.

\textbf{(e) Hardware Config:} Our experiments were run on a Compute Canada node having a Linux (CentOS 7) Operating System with 64GB primary memory (i.e., RAM) and 16GB GPU Memory (NVIDIA V100 Tensor Core GPU).

\subsubsection{Baseline Preparation}
To prepare the baselines, we utilized the replication packages provided by the authors of the original studies~\cite{fu2022linevul, hoang2019deepjit, he2020duplicate}. This allowed us to accurately reproduce the baseline setups and maintain the integrity of their original implementations. We prepared three variants of the original baselines for our analysis, which are described below.

\looseness=-1
\textbf{1. Baseline with Clean Data:} First, we obtained clean datasets from existing studies~\cite{croft2023data, zhang2023duplicate, issta2021deepjit} and prepared the original techniques to be trained using this high-quality data (Step \circled{4}, Fig.~\ref{fig:schematicDiagram}). This variant represents the ideal setup for the models as it is prepared with clean and high-quality data.

\textbf{2. Baseline with Buggy Data: }To investigate the impact of data quality on model performance, we created a second variant prepared with data quality issues. We obtained buggy datasets from the same existing studies~\cite{croft2023data, zhang2023duplicate, issta2021deepjit} and configured the baseline models to be trained using this buggy data (Step \circled{5}, Fig.~\ref{fig:schematicDiagram}). This step will allow us to observe how bugs in the training data might affect the model's learning process and subsequent performance.

\looseness=-1
\textbf{3. Baseline with Missing Preprocessing}: For our third variant, we focused on preprocessing faults, which are among the most prevalent types of faults in deep learning, according to the existing taxonomy~\cite{taxonomyRealFaults}. We removed the preprocessing operations from our training pipeline dealing with different types of data (Step \circled{6}, Fig.~\ref{fig:schematicDiagram}): for code-based data, we omitted the stop word removal, line separation, and sequence length normalization; for text-based data, we excluded stemming, stop word removal, and case conversion; and for metric-based data, we eliminated data normalization and feature scaling. By setting up the model to simulate the missing preprocessing operation, we aim to understand how the absence of these crucial steps might impact a model's performance and robustness.

\subsubsection{Integrating Weights and Biases}
To better understand the training process and model behaviour, we integrated Weights and Biases (W\&B) logging~\cite{wandb} into the existing baselines and their variants (Step \circled{7}, Fig.~\ref{fig:schematicDiagram}). W\&B's built-in logging capabilities and predefined hooks facilitated the extraction of gradients, weights, and biases. After configuring W\&B to watch the model parameters, it automatically tracked these metrics and attached hooks to each layer, eliminating the need for manual implementation. During training, W\&B captured the full distribution of gradients flowing through each layer, weight matrices, and bias vectors. The hooks tracked gradient changes during backpropagation, providing insights into the learning dynamics of different layers. These metrics were stored in W\&B's dashboard for detailed analysis of the model's internal dynamics throughout the training process. We implemented this integration across all model variants to maintain consistency in our experimental methodology. Additionally, we logged runtime configurations to document the computational environment. To account for the variability and stochasticity inherent in neural networks, we ran each baseline model and corresponding variants fifteen times\cite{arcuri2014hitchhiker}.

\subsection{Quantitative Analysis}
\looseness=-1
To quantitatively analyze the impact of data quality and preprocessing issues on the deep learning models, we leveraged the data from Weights and Biases (W\&B) captured during the training process. We use the gradients, weights, and biases over multiple runs and calculate aggregated statistics for each layer of the models (Step \circled{7}, Fig.~\ref{fig:schematicDiagram}). Table.~\ref{tab:weight_metrics} discusses the aggregate metrics used for analysis. By examining these aggregated statistics, we identified symptoms and manifestations of data quality and preprocessing issues and quantified their impact on deep learning models. We analyzed the statistical differences between models trained on clean datasets and those trained on buggy datasets to identify if there were any differences in their performance. We observed how data bugs and preprocessing issues affected different layers' weights, biases, and gradients. By quantifying the impact on specific parts of the models, such as attention layers, convolutional layers, and fully connected layers, we gained insights into how data quality and preprocessing issues can degrade the performance and reliability of deep learning models in software engineering tasks.

\subsection{Determining the Symptoms}
We developed a data-driven approach to identify symptoms of data quality and preprocessing issues in deep learning models. Our methodology consists of four key steps:

First, we established baseline parameter distributions by training 15 models on clean, properly preprocessed data. For each model layer, we collected descriptive statistics of weights, biases, and gradients during training. In particular, we capture the mean, median, maximum, minimum, variance, standard deviation, skewness, kurtosis and sparsity of these measures. These statistics characterize the expected behaviors of properly trained models and serve as our reference point.

Second, we trained multiple models on datasets containing known preprocessing and data quality issues. Similar to the first step, we captured descriptive statistics of weights, biases and gradients.

Third, we aggregated these statistics layer-by-layer across multiple models to identify common patterns. We then compared the aggregated statistics between the bug-free and buggy models to detect deviations in the buggy model's statistics, using the bug-free model as our reference. We also flagged outliers in the buggy model's statistics using three complementary statistical methods:
\begin{itemize}
\item Z-Score analysis for weights and biases: Computing mean ($\mu$) and standard deviation ($\sigma$) identify data points with high Z-score, $|Z|>3$ (i.e., $|X-\mu|>3\sigma$)~\cite{tukey1977exploratory}.
\item Interquartile Range (IQR) for gradients: Calculating 25th ($Q_1$) and 75th ($Q_3$) percentiles to identify values outside $[Q_1-1.5\times IQR, Q_3+1.5\times IQR]$~\cite{leys2013detecting}.
\item Distribution shape analysis: Measuring skewness ($>1$) and kurtosis \\ ($|kurtosis|>3$) to detect distribution distortions~\cite{westfall2014kurtosis}.
\end{itemize} 

These observed statistical deviations are the "symptoms" of data quality issues. The correlation between a symptom and a specific data quality issue is established by observing consistent patterns in our controlled experiments. For instance, when we introduce a bug like class imbalance, a specific statistical pattern (e.g., high gradient skewness) consistently emerges across multiple training runs. Since this pattern is absent in our bug-free baseline models, we can directly correlate the bug with its statistical symptom.

Finally, we analyzed statistics and averaged them across models and layers where these patterns appeared consistently in the buggy models. We then compared these against our bug-free models, which provided essential reference distributions of normal parameter behavior. These bug-free models were crucial for establishing reliable statistical thresholds and understanding what constitutes expected parameter distributions in properly trained models. Therefore, the thresholds are the specific statistical boundaries derived from this comparative analysis (e.g., a Z-score $>$ 3 or skewness $>$ 1), which signal a high probability of a data quality issue when exceeded. These thresholds were also validated against validation datasets to confirm their effectiveness in detecting known data quality issues. This process resulted in a set of reliable statistical criteria that can help identify symptoms and their prevalence in model training.

\subsection{Determining the Impact on Downstream Performance}
After determining the symptoms of data quality and preprocessing issues, we determine the impact of these symptoms on the performance of the downstream task. To determine how individual symptoms affect model performance while controlling for confounding factors; we performed a systematic analysis of models exhibiting isolated symptoms. For each symptom identified through our quantitative analysis, we filtered our experimental runs to obtain a subset of models that displayed only that particular symptom, excluding any models that exhibited multiple concurrent symptoms. This filtering process ensured that our analysis focused solely on the direct relationship between individual symptoms and performance degradation. We then aggregated performance metrics (accuracy, precision, recall, F1-score) across all models that uniquely exhibited each type of symptom. By analyzing models with isolated symptoms, we could attribute performance changes to specific symptoms without the confounding effects of symptom co-occurrence.

\subsection{Post-Hoc Analysis}
To gain deeper insights into the impacts of data bugs on our deep learning models, we conducted a post-hoc analysis using Explainable AI techniques (Step \circled{8}, Fig.~\ref{fig:schematicDiagram}). These techniques help us identify the influential parts of the input data and understand how the models make predictions.

For the code-based data, we analyzed LineVul's attention weights. Since the LineVul model is attention-based, an analysis of its attention weights helps us identify the parts of the input code that the model focuses on when making predictions about vulnerabilities. By comparing the attention weights from models trained on clean and buggy data, we observed how data bugs affect a model's attention and potentially lead to incorrect predictions. Given an attention weight tensor $A^l \in \mathbb{R}^{h \times n \times n}$ from layer $l$, which contains attention weights for $h$ heads and $n$ tokens, we first average the attention weights across all heads. For each token pair $(j,k)$ in the attention matrix, we compute the mean attention weight $\bar{A}^l_{jk}$ using:

\begin{equation}
\bar{A}^l_{jk} = \frac{1}{h}\sum_{i=1}^{h}A^l_{i,j,k}
\end{equation}

This gives us a single $n \times n$ attention matrix (for n tokens) that represents the average attention weights across all heads. To quantify the overall importance of each token $j$ in the sequence, we sum up all the attention weights directed towards that token using:

\begin{equation}
score_j = \sum_{i=1}^{n}\bar{A}^l_{ij}
\end{equation}

This produces a scalar $score_j$ for each token $j$, representing how much attention that token receives from all other tokens on average. When calculated for all tokens, this results in a vector of importance scores $[score_1, ..., score_n]$, which measures each token's relative importance in the context.

For the text-based data, we utilized t-SNE (t-Distributed Stochastic Neighbor Embedding) visualization~\cite{van2008visualizing}. t-SNE is a dimensionality reduction technique that helps us visualize high-dimensional data in a lower-dimensional space. By applying t-SNE to the representations learned by the DCCNN model~\cite{he2020duplicate} at different layers, we visualized the decision boundary for duplicate and non-duplicate bug reports. Given high-dimensional embeddings $X = {x_1, ..., x_N}$ and their low-dimensional counterparts $Y = {y_1, ..., y_N}$, t-SNE minimizes their distribution gap by employing the Kullback-Leibler divergence:

\begin{equation}
KL(P||Q) = \sum_{i}\sum_{j}p_{ij}\log\frac{p_{ij}}{q_{ij}}
\end{equation}

where $p_{ij}$ and $q_{ij}$ are the pairwise similarities of the duplicate bug reports in the high and low-dimensional spaces, respectively. In the high-dimensional space, similarities are computed using a Gaussian distribution:

\begin{equation}
p_{ij} = \frac{\exp(-||x_i-x_j||^2/2\sigma^2)}{\sum_{k\neq i}\sum_{l\neq k}\exp(-||x_k-x_l||^2/2\sigma^2)}
\end{equation}

In the low-dimensional space, similarities are computed using a Student's t-distribution with one degree of freedom:

\begin{equation}
q_{ij} = \frac{(1 + ||y_i-y_j||^2)^{-1}}{\sum_{k\neq i}\sum_{l\neq k}(1 + ||y_k-y_l||^2)^{-1}}
\end{equation}

The t-SNE based visualization above helps us understand how the learned representations differ between models trained on clean and buggy data, providing insights into the impact of data bugs on the model's learning process.

In the case of metric-based data, we implemented GradCAM~\cite{selvaraju2017grad} for our analysis. In the context of defect prediction using DeepJIT, GradCAM helps us visualize which metrics and parts of the input features are most influential in the model's decision, as per the existing literature~\cite{issta2021deepjit}. For a given class $c$, let $y^c$ be the score for class $c$ before softmax. The GradCAM weights $\alpha^c_k$ for each feature map $k$ are computed as:

\begin{equation}
\alpha^c_k = \frac{1}{Z}\sum_{i}\sum_{j}\frac{\partial y^c}{\partial A^k_{ij}}
\end{equation}

where $A^k$ represents the $k$-th feature map in the network, which corresponds to learned representations of different aspects of our input metrics. Unlike image applications where $(i,j)$ refers to spatial positions, in our metric-based data, $i$ indexes the samples (code changes) and $j$ indexes the feature dimensions (software metrics). The term $Z$ normalizes the gradients across all input instances.

The final GradCAM heatmap is obtained through a weighted combination of feature maps followed by a ReLU activation:

\begin{equation}
L^c_{GradCAM} = ReLU\left(\sum_k \alpha^c_k A^k\right)
\end{equation}

This equation performs a linear combination of the feature maps $A^k$, where each map is weighted by its importance coefficient $\alpha^c_k$. The ReLU function (Rectified Linear Unit) then sets all negative values to zero, ensuring that only features positively contributing to the class prediction are highlighted in the visualization. The resulting heatmap $L^c_{GradCAM}$ has the same dimensions as our feature matrix and indicates which metrics and feature combinations most strongly influence the model's classification decision for class $c$.

By contrasting the GradCAM outputs of models trained on clean data against those trained on buggy data, we identify how data bugs affect the model's attention to different metrics and potentially lead to incorrect defect predictions.

\begin{table}
\centering
\caption{Aggregate Metrics for Model Analysis}
\label{tab:weight_metrics}
\resizebox{\textwidth}{!}{
\begin{tabular}{|l|l|}
\hline
\textbf{Operator} & \textbf{Description} \\ \hline
max & The maximum value of the property in the layer. \\ \hline
min & The minimum value of the property in the layer. \\ \hline
median & The median value of the property in the layer. \\ \hline
mean & The average value of the property in the layer. \\ \hline
var & The variance of the property in the layer. \\ \hline
std & The standard deviation of the property in the layer. \\ \hline
skew & A measure of the asymmetry of the distribution of the property in the layer. \\ \hline
kurt & A measure of the peakedness and tail heaviness of the property's distribution in the layer. \\ \hline
spar & The fraction of properties in a layer that are zero or close to zero. \\ \hline
\end{tabular}
}
\end{table}

\subsection{Validating the Derived Findings}
To assess the generalizability of our findings from RQ1, RQ2, and RQ3, we conducted additional experiments using six new datasets not included in our initial analysis (Step \circled{9}, Fig.~\ref{fig:schematicDiagram}), following similar approaches in existing studies~\cite{hossain2024deep}. In particular, we chose six new datasets representing three types of data: two code-based datasets (Juliet and D2A), two text-based datasets (Mozilla and Spark), and two metric-based datasets (Go and JDT). We collect the buggy and clean versions of these datasets from the same studies~\cite{zhang2023duplicate, issta2021deepjit, croft2023data}, which were used in our previous steps. Table ~\ref{tab:validationDatasets} summarises our validation datasets, including their sizes, compositions, and brief descriptions.

\begin{table}[h]
\centering
\caption{Summary of datasets used for validation and generalization}
\label{tab:validationDatasets}
\resizebox{\textwidth}{!}{%
\begin{tabular}{|l|l|l|l|}
\hline
\textbf{Data Type} & \textbf{Dataset} & \textbf{Size} & \textbf{Description} \\ \hline
\multirow{2}{*}{Code-Based} &
Juliet &
253,002 test cases &
C/C++ and Java test cases covering 181 CWEs \\ \cline{2-4}
& D2A &
1,295,623 samples &
Samples from six open-source projects such as OpenSSL, FFmpeg etc.\\ \hline
\multirow{2}{*}{Text-Based} &
Mozilla &
193,587 bug reports &
Bug reports from Mozilla projects \\ \cline{2-4}
& Spark &
9,579 bug reports &
Bug reports from Apache Spark project \\ \hline
\multirow{2}{*}{Metric-Based} &
Go &
61,224 files &
Metrics from Go programming language project \\ \cline{2-4}
& JDT &
13,348 files &
Metrics from Eclipse Java Development Tools \\ \hline
\end{tabular}%
}
\end{table}

In our validation, we retrained each of the baseline models (LineVul, DCCNN, and DeepJIT) with their corresponding datasets containing clean and buggy data. We performed similar quantitative analyses, including monitoring training behaviour and examining model components. Moreover, we also perform the qualitative analyses using the same post-hoc techniques (attention weight analysis, t-SNE visualization, and GradCAM) as in our original study. By comparing the results from these validation experiments with our initial findings, we aimed to determine whether the observed impacts of data quality issues are consistent across different datasets within each data type, and this assessed the generalizability of our conclusions.

\section {Study Findings}
In our study, we examined the effects of four common data bugs: label noise, class imbalance, concept drift, and missing preprocessing. To address each research question, we constructed and trained a comprehensive dataset of 120 buggy models, with 30 models dedicated to each bug type. Additionally, we trained 30 models on clean data to serve as a baseline, helping us establish the expected training behaviour. We used the W\&B logging framework~\cite{wandb} to capture the training behaviours of both faulty and clean models. This section presents our findings, focusing on the most common symptoms of these data bugs across three data types: code-based, text-based, and metric-based.

\subsection{RQ1: How do data quality and preprocessing issues in code-based data affect the training behaviour of deep learning models?}

\subsubsection{Impact of Data Quality}
\looseness=-1
\textbf{(a) Near-Zero Biases}: When we trained our models using buggy data, we observed near-zero biases in several layers. Normal learning is characterized by well-defined bias values (e.g., $\geq$0.5), while near-zero biases (e.g., $<$ 0.01) might suggest learning difficulties~\cite{neal2018modern}. The bias parameter introduces non-linearity in the network and contributes to the learning process~\cite{zhang2024bias, williams2024expressivity}. Thus, near-zero bias values might indicate a lack of non-linearity in the neural network. As shown in Table~\ref{tab:rq1Table}, 76.67\% of the models trained on data with label noise and 50.00\% of the models trained on data with concept drift demonstrated near-zero biases. Our manual analysis shows that $\approx$20\% of their layers were affected, especially the attention and output layers, which exhibited near-zero biases ($<$ 0.01) due to data bugs. In contrast, unaffected layers (e.g., embedding layer) maintained normal bias values ($\geq$ 0.5) and had regular parameter updates during backpropagation. These disparities in bias values across different layers indicate inconsistency in the learning capabilities of the layers~\cite{ye2017importance}. We also observed performance deterioration in the models with near-zero biases. As shown in Table~\ref{tab:codeMetricImpact}, LineVul witnessed an 8.12\%-8.39\% decline across various performance metrics. CodeBERT exhibited similar behaviours for data bugs, with metrics decreasing by $\approx$6\%.

\textbf{(b) Smaller Weights}: We observed small weights in our models trained on buggy data. According to existing studies, small and tightly clustered weights in a neural network indicate reduced representational power of the neurons~\cite{liu-etal-2020-understanding}. As shown in Table~\ref{tab:rq1Table}, 83.33\% of models trained on data with label noise and 56.67\% of models trained with concept drift demonstrated a symptom of very small weights. Our analysis of these models revealed that 90-95\% of their layers were affected, particularly embedding, attention, and dense layers. These layers exhibited small weights (e.g., $\mu\approx0.011$, $\sigma \approx 0.009$ and $\mu\approx0.047$, $\sigma \approx 0.038$), significantly lower than those observed in models trained on clean data (e.g., $\mu\approx0.23$, $\sigma \approx 0.11$). These small weights had a significant impact on the model's performance, causing a 14.77\%-16.48\% and 6.82\%-13.17\% reduction in various evaluation metrics for our baseline techniques (LineVul, CodeBERT).

\textbf{(c) Gradient Instability}: When we trained our models using buggy data, we observed instability in their gradients. From the perspective of neural networks, gradient instability indicates extreme variations in model parameter updates, which prevents effective signal propagation during the backpropagation step~\cite{storm2020unstable}. Moreover, existing literature has shown that the extreme gradient values might lead to exploding gradients and vanishing gradients~\cite{pascanu2013difficulty}. We found that models trained on buggy data exhibited significant gradient instability. As shown in Table~\ref{tab:rq1Table}, 68.33\% of models trained on data with label noise and 53.33\% with class imbalance demonstrated this issue. Our analysis revealed that 30-40\% of their layers were affected, especially the attention and dense layers. We found that these layers have a gradient range between $10^{-9}$ and $10^{1}$, which is significantly larger than the gradient range from models trained on clean data — ($10^{-3}$ to $10^{1}$). These gradient instabilities also had a notable impact on the model's performance, causing an 11.75\%-13.01\% and 10.01\%-11.16\% reduction in various performance metrics for our two baseline techniques (LineVul, CodeBERT).
 
\begin{table}
\centering
\setlength{\tabcolsep}{4pt}
\caption{Manifestations of Data Bugs in Code-Based Models}
\label{tab:rq1Table}
\resizebox{\columnwidth}{!}{%
\begin{tabular}{|l|l|l|l|}
\hline
\textbf{Data Quality Issues} & \textbf{Near-Zero Biases} & \textbf{Smaller Weights} & \textbf{Gradient Instability} \\ \hline
\multicolumn{4}{|c|}{\textbf{LineVul}} \\ \hline
Label Noise    & 80.00\% & 86.67\% & 73.33\% \\ \hline
Class Imbalance & 6.67\% & 23.33\% & 53.33\% \\ \hline
Concept Drift   & 56.67\% & 63.33\% & 16.67\% \\ \hline
\multicolumn{4}{|c|}{\textbf{CodeBERT}} \\ \hline
Label Noise    & 73.33\% & 80.00\% & 63.33\% \\ \hline
Class Imbalance & 50.00\% & 60.00\% & 53.33\% \\ \hline
Concept Drift   & 43.33\% & 50.00\% & 26.67\% \\ \hline
\end{tabular}
}
\end{table}

\begin{table}
\centering
\setlength{\tabcolsep}{4pt}
\caption{Impact of Data Quality Issues on Model's Performance for Code-Based Data}
\label{tab:codeMetricImpact}
\resizebox{\columnwidth}{!}{%
\begin{tabular}{|l|l|l|l|l|}
\hline
\textbf{Symptoms} & \textbf{Accuracy} & \textbf{Precision} & \textbf{Recall} & \textbf{F1-Score} \\ \hline
\multicolumn{5}{|c|}{\textbf{LineVul}} \\ \hline
Performance on Clean Data & 93.62\% & 94.24\% & 83.12\% & 88.45\% \\ \hline
Near-Zero Biases & 85.23\% & 86.12\% & 74.89\% & 80.13\% \\ \hline
Smaller Weights & 77.14\% & 78.92\% & 68.35\% & 73.24\% \\ \hline
Gradient Instability & 81.87\% & 81.23\% & 71.96\% & 76.31\% \\ \hline
\multicolumn{5}{|c|}{\textbf{CodeBERT}} \\ \hline
Performance on Clean Data & 67.25\% & 61.25\% & 58.63\% & 55.25\% \\ \hline
Near-Zero Biases & 59.87\% & 56.12\% & 50.23\% & 52.94\% \\ \hline
Smaller Weights & 54.13\% & 49.26\% & 47.89\% & 48.43\% \\ \hline
Gradient Instability & 57.24\% & 52.87\% & 49.12\% & 50.86\% \\ \hline
\end{tabular}
}
\end{table}

\subsubsection{Impact of Missing Preprocessing Operations}

\textbf{(a) Slow Convergence}: When we trained our models without any appropriate preprocessing operations, they converged slowly. That is, these models took longer than expected to reach optimal parameter values during training. To identify this issue, we analyze the time (i.e., epochs) required for convergence.  Out of all the studied preprocessing operations for code-based data, missing line separation had the most severe impact, affecting 70.00\% of LineVul and 76.67\% of CodeBERT models. The models trained with proper preprocessing operations converged to the optimal parameters in $\approx$3.7 epochs, whereas the models without such preprocessing could not converge even after 5 epochs. Furthermore, the impact of slow convergence on the model's performance was evident, causing a 10.01\%-11.14\% and 3.11\%-8.40\% reduction in various performance metrics for our two baseline techniques (LineVul, CodeBERT).

\textbf{(b) Extreme Bias Values}: When we trained the models without proper preprocessing, we observed a high range in their bias values. The absence of stop word removal had the highest impact, affecting 56.67\% of LineVul and 63.33\% of CodeBERT models. Our analysis revealed that $\approx$60\% of their layers were affected, showing widely dispersed bias values ranging between 0.03 and 1.97, significantly different from those of error-free models: (e.g., 0.34 to 0.72). Contained bias values indicate stable parameter distribution and learning patterns during model training~\cite{williams2024expressivity}. We also found that models with extreme bias values suffer from a substantial performance degradation, causing a 13.77\%-17.32\% and 7.72\%-13.12\% reduction in various performance metrics for our two baseline techniques (LineVul, CodeBERT).

\textbf{(c) Skewed Parameter Distributions}: When we trained our models without appropriate preprocessing, we also observed skewed weight distributions across the model's layers. The absence of sequence length normalization had the highest impact, affecting 73.33\% of LineVul and 83.33\% of CodeBERT models. Our analysis revealed that 80\% of their layers were affected, particularly embedding, attention, and dense layers. In models trained on clean code data, weights exhibited moderate skewness (0.4) and kurtosis (1.3), indicating balanced distributions. However, models trained without preprocessing showed high skewness ($>1$) and extreme kurtosis ($|k| \gg 3$), deviating significantly from clean data distributions. We also noticed a performance degradation in the affected models, causing a 10.16\%-13.01\% and 5.39\%-10.51\% reduction in various performance metrics for our two baseline techniques (LineVul, CodeBERT).

\begin{table}[h]
\centering
\setlength{\tabcolsep}{4pt}
\caption{Manifestations of Missing Preprocessing in Code-Based Models}
\label{tab:rq1PreprocessTable}
\resizebox{\columnwidth}{!}{%
\begin{tabular}{|l|l|l|l|}
\hline
\textbf{Preprocessing Issues} & \textbf{Slow Convergence} & \textbf{Extreme Bias Values} & \textbf{Skewed Parameter Distributions} \\ \hline
\multicolumn{4}{|c|}{\textbf{LineVul}} \\ \hline
Stop Word Removal & 68.33\% & 56.67\% & 70.00\% \\ \hline
Line Separation & 70.00\% & 50.00\% & 66.67\% \\ \hline
Sequence Length Normalization & 63.33\% & 53.33\% & 73.33\% \\ \hline
\multicolumn{4}{|c|}{\textbf{CodeBERT}} \\ \hline
Stop Word Removal & 73.33\% & 63.33\% & 76.67\% \\ \hline
Line Separation & 76.67\% & 60.00\% & 70.00\% \\ \hline
Sequence Length Normalization & 70.00\% & 56.67\% & 83.33\% \\ \hline
\end{tabular}
}
\end{table}

\begin{table}
\centering
\setlength{\tabcolsep}{4pt}
\caption{Impact of Missing Preprocessing on Model's Performance for Code-Based Data}
\label{tab:rq1preprocessMetricImpact}
\resizebox{\columnwidth}{!}{%
\begin{tabular}{|l|l|l|l|l|}
\hline
\textbf{Symptoms} & \textbf{Accuracy} & \textbf{Precision} & \textbf{Recall} & \textbf{F1-Score} \\ \hline
\multicolumn{5}{|c|}{\textbf{LineVul}} \\ \hline
Performance on Clean Data & 93.62\% & 94.24\% & 83.12\% & 88.45\% \\ \hline
Slow Convergence & 83.14\% & 84.23\% & 71.98\% & 77.62\% \\ \hline
Extreme Bias Values & 77.14\% & 76.92\% & 69.35\% & 72.89\% \\ \hline
Skewed Parameter Distributions & 81.87\% & 81.23\% & 72.96\% & 76.85\% \\ \hline
\multicolumn{5}{|c|}{\textbf{CodeBERT}} \\ \hline
Performance on Clean Data & 67.25\% & 61.25\% & 58.63\% & 55.25\% \\ \hline
Slow Convergence & 58.87\% & 54.12\% & 50.23\% & 52.14\% \\ \hline
Extreme Bias Values & 54.13\% & 48.26\% & 46.89\% & 47.53\% \\ \hline
Skewed Parameter Distributions & 57.24\% & 51.87\% & 48.12\% & 49.86\% \\ \hline
\end{tabular}
}
\end{table}

\subsubsection{Impact analysis of Label Noise in Training Data using Attention Mechanisms}
\looseness=-1
Attention weights have been used in software engineering research to analyze the correlation between the input and the output of attention-based models~\cite{cao2024systematic, wang2020detecting, ma2024unveiling}. In our study, we also analyze the attention weights of our model to understand the impact of data quality issues on the model's training. Among four data quality issues, we found that label noise had the strongest impact on models trained with code-based data (see Table~\ref{tab:rq1Table}). Thus, we conduct a qualitative analysis to better understand how label noise affects a model during training. Towards this, we trained 60 models in total. We divided these models into four groups of 15 models each: (a) models trained on a clean Devign dataset, (b) models trained on a clean BigVul dataset, (c) models trained on a Devign dataset with label noise, and (d) models trained on a BigVul dataset with label noise. For each group, we calculated the average attention weights for the tokens in the BigVul dataset across all 15 models. By analyzing 15 models from each group, we aimed to mitigate the impact of non-determinism in deep learning. We discuss our findings below.

\textbf{Bug-free Data:} Models trained on the bug-free versions of our datasets demonstrate a strong focus on the tokens related to security flaws. BigVul dataset contains source code from Chrome, Linux, Android, and Tcpdump. When analyzing models trained on BigVul, we observe high attention weights for key functions and data structures related to security vulnerabilities. For instance, \texttt{rds6\_inc\_info\_copy} (87.3796) and \texttt{struct rds\_incoming} (64.1523) receive significant attention, indicating their relevance to network-related vulnerabilities. The models also heavily emphasize network fields such as \texttt{laddr} (85.7431) and \texttt{faddr} (86.2221), demonstrating their awareness of potential security issues in network communications. The Devign dataset contains source code from Linux, FFmpeg, Qemu, and Wireshark. Models trained on this dataset focus on memory-related operations and type conversions, major causes of security vulnerabilities~\cite{Stevens_Fenner_Rudoff_2013}. Tokens like \texttt{uint16\_t} (88.8645) and \texttt{uint64\_t} (86.9365) receive very high attention weights, which reflects their importance to data handling. Critical function calls for cross-platform compatibility, such as \texttt{le16\_to\_cpu} (82.2068) and \texttt{le64\_to\_cpu} (80.8885), are also strongly emphasized. Models trained on BigVul and Devign separately show significant attention to error-checking patterns and memory operations like \texttt{sizeof} (30.5034) and \texttt{if(!conn)} (32.1079). These attention weights suggest the models have developed a comprehensive understanding of code elements related to security flaws when the datasets do not contain any label noise.

\begin{figure}[htbp]
    \centering
    \begin{subfigure}{\linewidth}
        \includegraphics[width=1\linewidth]{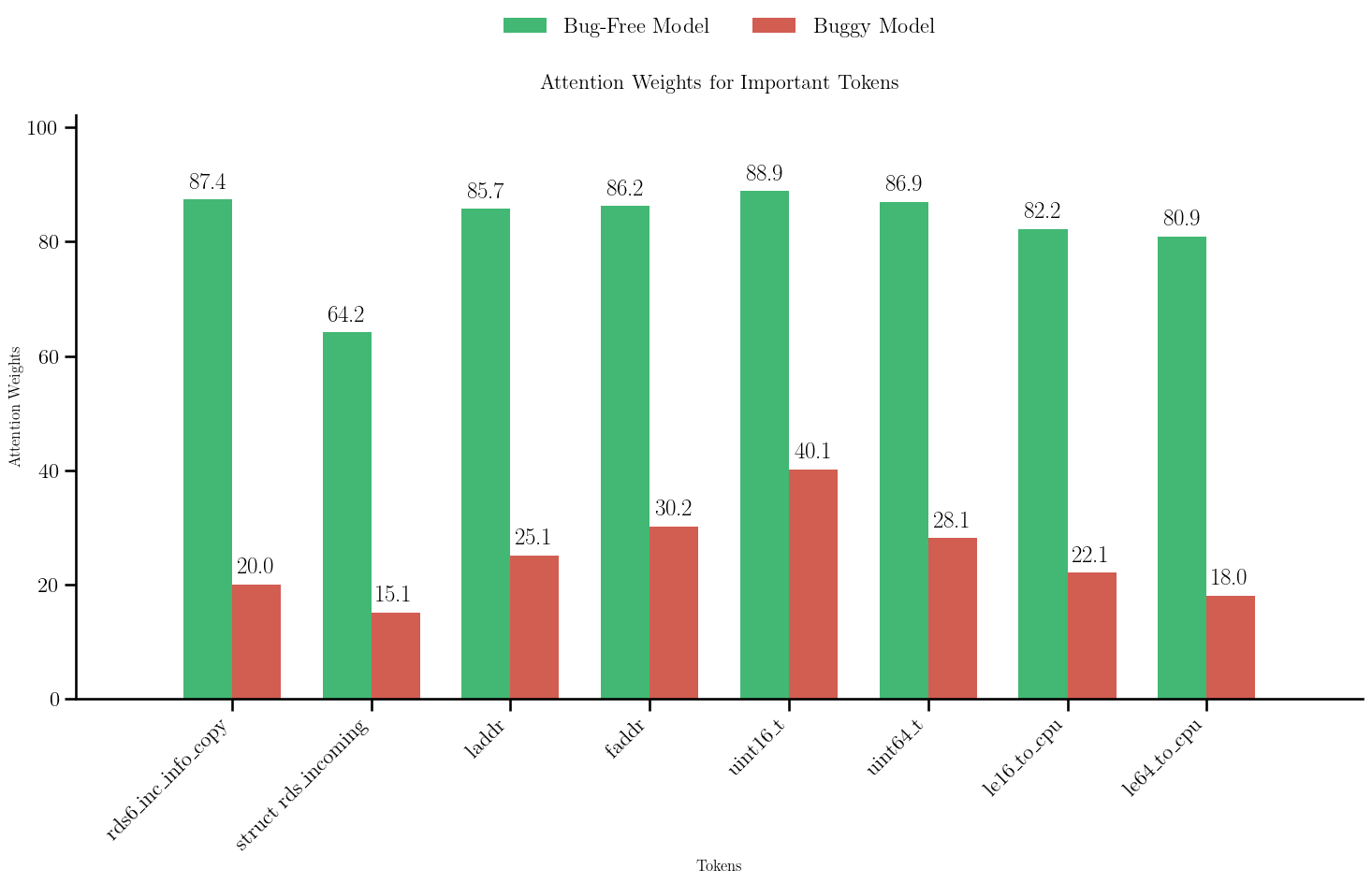}
        \caption{Attention weights assigned to salient tokens (i.e., related to security vulnerability) in the BigVul dataset by bug-free and buggy models.}
        \label{fig:important-tokens}
    \end{subfigure}
    \begin{subfigure}{\linewidth}
        \includegraphics[width=1\linewidth]{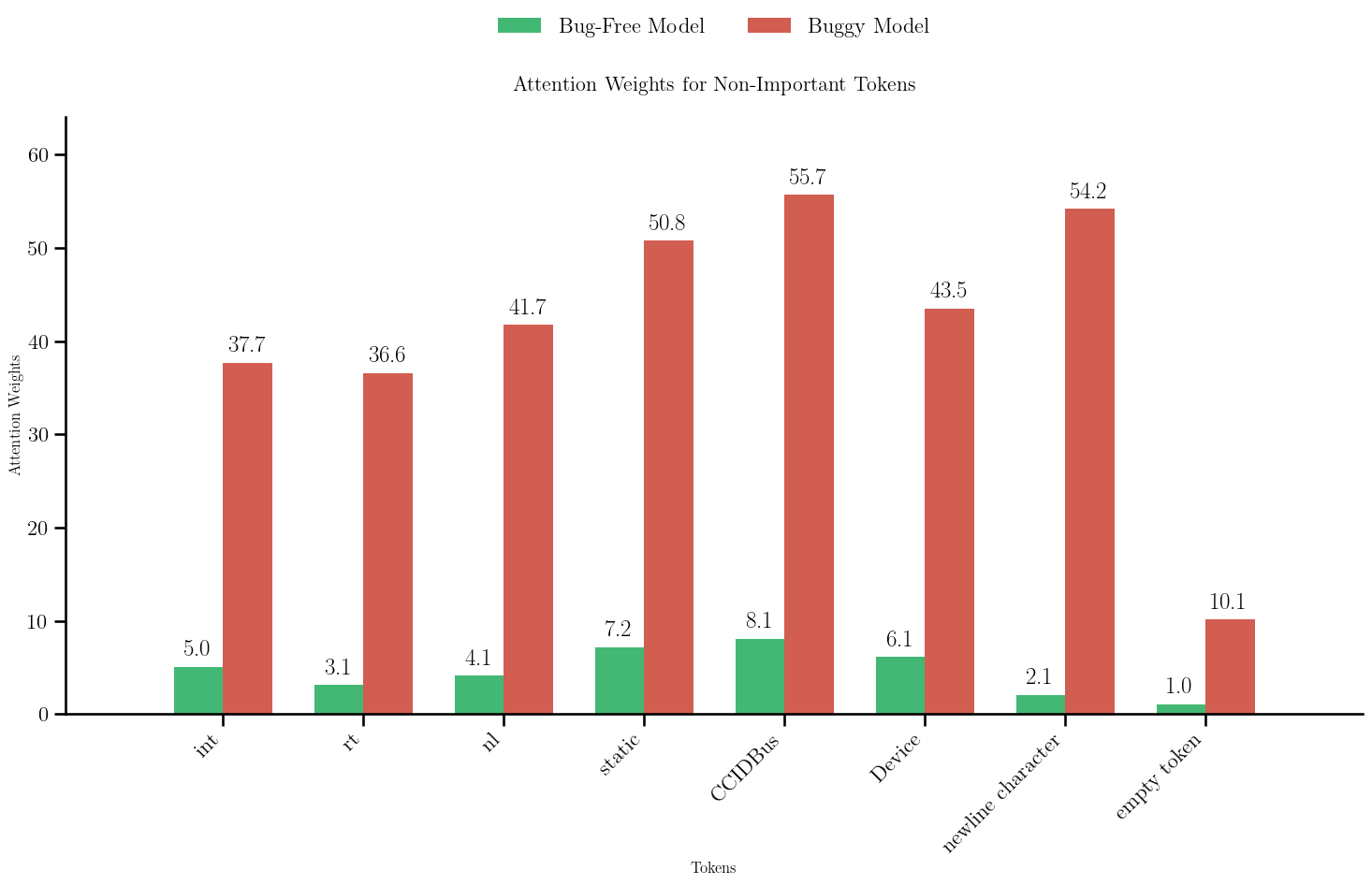}
        \caption{Attention weights assigned to non-vulnerable tokens in the BigVul dataset by bug-free and buggy models.}
        \label{fig:non-important-tokens}
    \end{subfigure}
    \caption{Comparison of attention weights between bug-free and buggy models for both important and non-important tokens. The bug-free model (green) shows higher attention weights for important tokens and lower weights for non-important tokens, while the buggy model (red) shows the opposite pattern.}
    \label{fig:attention-weights}
\end{figure}

\textbf{Buggy Data:} In contrast, our models trained on the datasets containing label noise showed their inability to focus on appropriate tokens. For example, our models trained on BigVul (with label noise) assign significant weights to such tokens that might not be related to security vulnerability: \texttt{int} (37.6882), \texttt{rt} (36.5832), and \texttt{nl} (41.7406). Similarly, the models trained on LineVul (with label noise) focus their attention on various elements like \texttt{static} (50.7911), \texttt{CCIDBus} (55.7003), and \texttt{Device} (43.4922). In other words, the models fail to prioritize the tokens that might signal potential vulnerabilities, such as unsafe operations or improper error handling. Moreover, these models assign big weights to non-informative tokens, such as newline characters (26.1465 in BigVul, 54.1878 in Devign) and empty tokens (8.2194 in BigVul, 10.1193 in Devign). This diffused attention pattern, evident across both datasets, suggests the models have difficulty separating safe code from potentially vulnerable code. This limits their effectiveness in identifying security-critical code sections.

We also compare the attention values for individual tokens across the models. Figure~\ref{fig:attention-weights} shows fundamentally different attention patterns between bug-free and buggy models. The bug-free models trained on the BigVul dataset exhibit strong discrimination power by assigning high weights (80-89) to security-critical tokens like \texttt{uint16\_t} (88.9) and network identifiers (\texttt{laddr}, \texttt{faddr}) (85-87), while maintaining minimal attention (1-8) on non-essential elements. In stark contrast, the buggy models pay little attention (15-40) to security-critical tokens while heavily focusing on basic syntax elements and identifiers - \texttt{CCIDBus} (55.7), \texttt{static} (50.8), and newline characters (54.2). Given the weights to appropriate tokens, bug-free models have a higher chance of differentiating between vulnerable and benign code, where the model with a noisy dataset could fall short. Our experimental results (e.g., Table~\ref{tab:codeMetricImpact}) also show strong evidence to support this.

\begin{rqbox}
\textbf{Summary of RQ1:} Models trained on code-based datasets containing data bugs exhibit many issues, including near-zero biases, smaller weights, and gradient instability. Furthermore, missing preprocessing can lead to slow convergence, extreme bias values, and skewed parameter distributions, impairing the models' ability to understand and process code effectively.
\end{rqbox}

\subsection{RQ2: How do data quality and preprocessing issues in text-based data affect the training behaviour of deep learning models?}

\subsubsection{Impact of Data Quality}
\looseness=-1
\textbf{(a) Abnormal Weight Distribution}: When we trained our models using buggy text data, we observed abnormal weight distributions across multiple layers. According to existing studies, abnormal weight distributions can reduce a model's effectiveness in capturing important features~\cite{liu-etal-2020-understanding, ye2017importance}. As shown in Table~\ref{tab:rq2Table}, 81.67\% of the models trained on data with concept drift demonstrated abnormal weight distributions. Our manual analysis shows that 80-85\% of their layers were affected, especially convolutional and dense layers. The models trained on the buggy text exhibited high weight variances (e.g., $\sigma^2 \approx 2.3$) significantly deviating from baseline values (e.g., $\sigma^2 \approx 0.64$). We also observed notable performance deterioration in models with abnormal weight distributions. As shown in Table~\ref{tab:textMetricImpact}, DCCNN witnessed a 5.18\%-7.28\% decline across various performance metrics. CodeBERT+MLP exhibited similar behavior, with performance metrics decreasing by 3.50\%-4.10\%.

\textbf{(b) Gradient Skewness}: We also observed significant gradient skewness in our models trained on buggy data. Gradient skewness indicates an asymmetry in the gradient distribution curve, with a heavy tail on any side. According to existing studies, asymmetric gradient distributions can lead to exploding and vanishing gradients, causing unstable parameter updates~\cite{pascanu2013difficulty}. As shown in Table~\ref{tab:rq2Table}, 68.33\% of models trained on data with concept drift and 48.33\% with label noise demonstrated gradient skewness. Our analysis of these models revealed that 80-90\% of their layers were affected, particularly convolutional and batch normalization layers. These layers had the gradients with significantly higher skewness values (e.g., $\gamma \approx 2.9$ for concept drift), significantly higher than those observed in models trained on clean data (e.g., $\gamma \approx 0.2$). As shown in Table~\ref{tab:textMetricImpact}, these skewed gradients also had a substantial impact on the model's performance, causing a 7.49\%-9.44\% and 5.66\%-6.50\% decline in various performance metrics for our baseline techniques (DCCNN, CodeBERT+MLP).

\textbf{(c) Overfitting}: When we trained our models using buggy text data, we observed overfitting, which can be characterized by unusually high bias values. From the perspective of neural networks, overfitting occurs when models memorize training data patterns instead of learning the patterns~\cite{williams2024expressivity, zhang2024bias}. Unlike weights, bias terms are added directly to the layer outputs without normalization. When these bias values are abnormally high (e.g., $med \approx 1.1$ for models trained on buggy data versus the baseline models' values of $med \approx 0.07$), they effectively dominate the computation, and this reduces the impact of the weights and inputs on the final prediction. Because of such high bias values, the neurons consistently produce similar activation values irrespective of the inputs, which causes the model to essentially memorize fixed responses. We found that models trained on imbalanced data exhibited significant overfitting in terms of extreme bias values. As shown in Table~\ref{tab:rq2Table}, 85.00\% of models trained with class imbalance demonstrated this issue. Our analysis revealed that 60-65\% of their layers were affected, especially the dense layers. We found that these layers have bias values around $med \approx 1.1$, significantly larger than the bias values from the baseline models -- $med \approx 0.07$. Overfitting also had a substantial impact on the model's performance, as DCCNN's and CodeBERT+MLP's performance metrics dropped by 8.05\%-9.26\% and 5.50\%-7.95\%, respectively (Table~\ref{tab:textMetricImpact}).

\begin{table}
\centering
\caption{Manifestations of Data Bugs in Text-Based Models}
\label{tab:rq2Table}
\resizebox{\columnwidth}{!}{%
\begin{tabular}{|l|l|l|l|}
\hline
\textbf{Data Quality Issues} & \textbf{Abnormal Weight Distribution} & \textbf{Gradient Skewness} & \textbf{Overfitting} \\ \hline
\multicolumn{4}{|c|}{\textbf{DCCNN}} \\ \hline
Label Noise & 16.67\% & 56.67\% & 3.33\% \\ \hline
Class Imbalance & 36.67\% & 23.33\% & 86.66\% \\\hline
Concept Drift & 93.33\% & 76.67\% & 43.66\% \\ \hline
\multicolumn{4}{|c|}{\textbf{CodeBERT + MLP}} \\ \hline
Label Noise & 10.00\% & 40.00\% & 3.33\% \\ \hline
Class Imbalance & 43.33\% & 16.67\% & 83.33\% \\ \hline
Concept Drift & 70.00\% & 60.00\% & 33.33\% \\ \hline
\end{tabular}
}
\end{table}

\begin{table}
\centering
\caption{Impact of Data Bug Symptoms on Model's Performance for Text-Based Data}
\label{tab:textMetricImpact}
\resizebox{\columnwidth}{!}{%
\begin{tabular}{|l|l|l|l|l|}
\hline
\textbf{Symptoms} & \textbf{Accuracy} & \textbf{Precision} & \textbf{Recall} & \textbf{F1-Score} \\ \hline
\multicolumn{5}{|c|}{\textbf{DCCNN}} \\ \hline
Baseline Performance & 93.41\% & 91.68\% & 92.45\% & 92.11\% \\ \hline
Abnormal Weight Distribution & 88.23\% & 86.94\% & 85.17\% & 86.04\% \\ \hline
Gradient Skewness & 85.92\% & 84.13\% & 81.26\% & 82.67\% \\ \hline
Overfitting & 84.15\% & 82.31\% & 85.87\% & 84.06\% \\ \hline
\multicolumn{5}{|c|}{\textbf{CodeBERT + MLP}} \\ \hline
Baseline Performance & 89.62\% & 82.23\% & 88.97\% & 85.14\% \\ \hline
Abnormal Weight Distribution & 86.12\% & 79.23\% & 84.87\% & 81.94\% \\ \hline
Gradient Skewness & 83.96\% & 77.14\% & 82.31\% & 79.63\% \\ \hline
Overfitting & 82.17\% & 74.28\% & 85.92\% & 79.64\% \\ \hline
\end{tabular}
}
\end{table}

\subsubsection{Impact of Missing Preprocessing Operations}
\looseness=-1
\textbf{(a) Extreme Weights}: When we trained our models without appropriate preprocessing operations, we observed extreme weight values across multiple layers. As shown in Table~\ref{tab:rq2PreprocessTableBaselines}, 66.67\% of DCCNN and 73.33\% of CodeBERT+MLP models trained without stop word removal demonstrated extreme weights. Our manual analysis shows that 80-90\% of their layers were affected, especially embedding and convolutional layers. The models trained without appropriate preprocessing operations exhibited extreme weight ranges (e.g., -2.16 to 2.54), significantly deviating from the stable values in baseline models (e.g., -0.1 to 0.1). These extreme weights also had a notable impact on the performance of the models. As shown in Table~\ref{tab:rq2preprocessTextMetricImpact}, DCCNN witnessed a 9.18\%-9.47\% decline across various performance metrics. CodeBERT+MLP exhibited similar behavior, with performance metrics metrics decreasing by 8.38\%-8.74\%.

\textbf{(b) Skewed Bias Distributions}: We also observed significant skewness in the bias distribution of our models trained without appropriate preprocessing. As shown in Table~\ref{tab:rq2PreprocessTableBaselines}, 53.33\% of DCCNN and 60.00\% of CodeBERT+MLP models trained without stop word removal demonstrated skewed bias distributions. Our analysis of these models revealed that 50-55\% of their layers were affected, particularly the convolutional layers. These layers had highly skewed bias values (e.g., $>1$) with moderate kurtosis (e.g., $|k|>3$), significantly different from the balanced distributions observed in models trained on properly preprocessed data (skewness: 0.34, kurtosis: 1.82). These skewed bias distributions also had a substantial impact on the performance of the models, causing 6.49\%-7.17\% and 5.75\%-6.47\% performance drops in various metrics (DCCNN, CodeBERT+MLP).

\begin{table}[h]
\centering
\caption{Manifestations of Missing Preprocessing in Text-Based Models}
\label{tab:rq2PreprocessTableBaselines}
\resizebox{\columnwidth}{!}{%
\begin{tabular}{|l|l|l|}
\hline
\textbf{Missing Preprocessing Operation} & \textbf{Extreme Weights} & \textbf{Skewed Bias} \\ \hline
\multicolumn{3}{|c|}{\textbf{DCCNN}} \\ \hline
Stemming & 63.33\% & 50.00\% \\ \hline
Stop Word Removal & 66.67\% & 53.33\% \\ \hline
Case Conversion & 60.00\% & 46.67\% \\ \hline
\multicolumn{3}{|c|}{\textbf{CodeBERT + MLP}} \\ \hline
Stemming & 70.00\% & 56.67\% \\ \hline
Stop Word Removal & 73.33\% & 60.00\% \\ \hline
Case Conversion & 66.67\% & 53.33\% \\ \hline
\end{tabular}%
}
\end{table}

\begin{table}
\centering
\caption{Impact of Missing Preprocessing on Model's Performance for Text-Based Data}
\label{tab:rq2preprocessTextMetricImpact}
\resizebox{\columnwidth}{!}{%
\begin{tabular}{|l|l|l|l|l|}
\hline
\textbf{Symptoms} & \textbf{Accuracy} & \textbf{Precision} & \textbf{Recall} & \textbf{F1-Score} \\ \hline
\multicolumn{5}{|c|}{\textbf{DCCNN}} \\ \hline
Performance on Clean Data & 93.41\% & 91.68\% & 92.45\% & 92.11\% \\ \hline
Extreme Weights & 84.23\% & 82.14\% & 83.98\% & 83.05\% \\ \hline
Skewed Bias Distributions & 86.92\% & 84.73\% & 85.16\% & 84.94\% \\ \hline
\multicolumn{5}{|c|}{\textbf{CodeBERT + MLP}} \\ \hline
Performance on Clean Data & 89.62\% & 82.23\% & 88.97\% & 85.14\% \\ \hline
Extreme Weights & 81.24\% & 75.16\% & 80.23\% & 77.61\% \\ \hline
Skewed Bias Distributions & 83.87\% & 76.92\% & 82.14\% & 79.44\% \\ \hline
\end{tabular}
}
\end{table}

\subsubsection{Qualitative Analysis: Impact of Concept Drift in Training Data on Decision Boundary}
\begin{figure}
\centering
\subcaptionbox{First dense layer}[.40\linewidth]{%
\includegraphics[width=\linewidth]{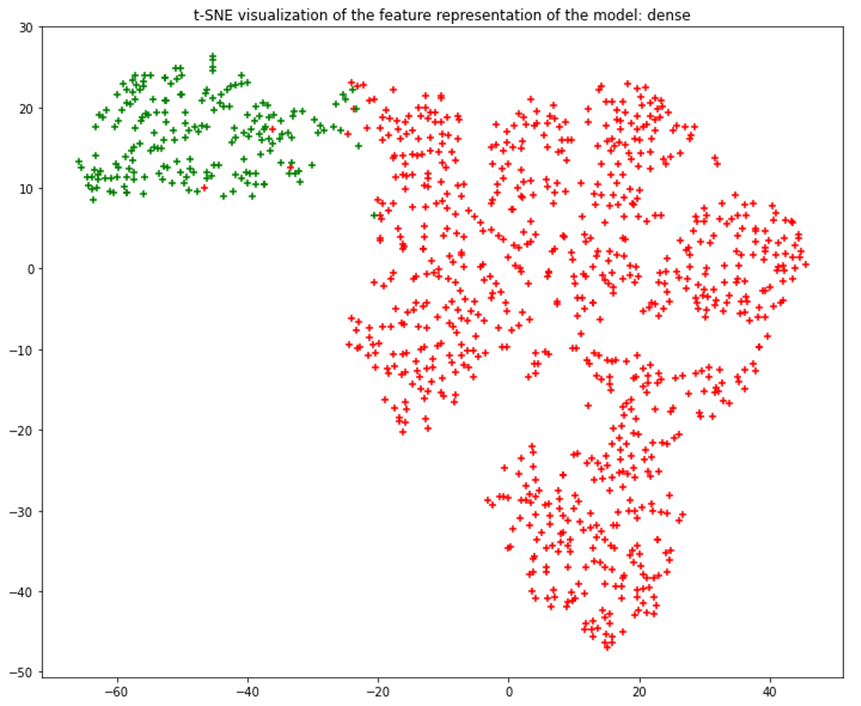}%
}%
\subcaptionbox{Second dense layer}[.40\linewidth]{%
\includegraphics[width=\linewidth]{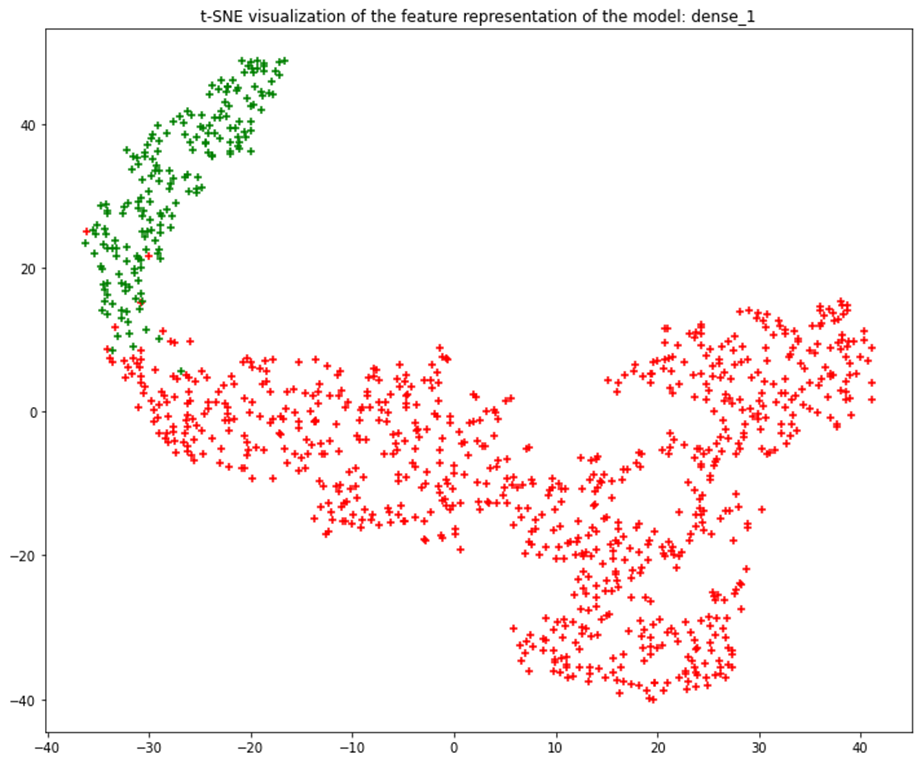}%
}
\caption{t-SNE plots for models trained on bug-free data}
\label{fig:bugfreeDense}
\end{figure}

\begin{figure}
\centering
\subcaptionbox{First dense layer}[.40\linewidth]{%
\includegraphics[width=\linewidth]{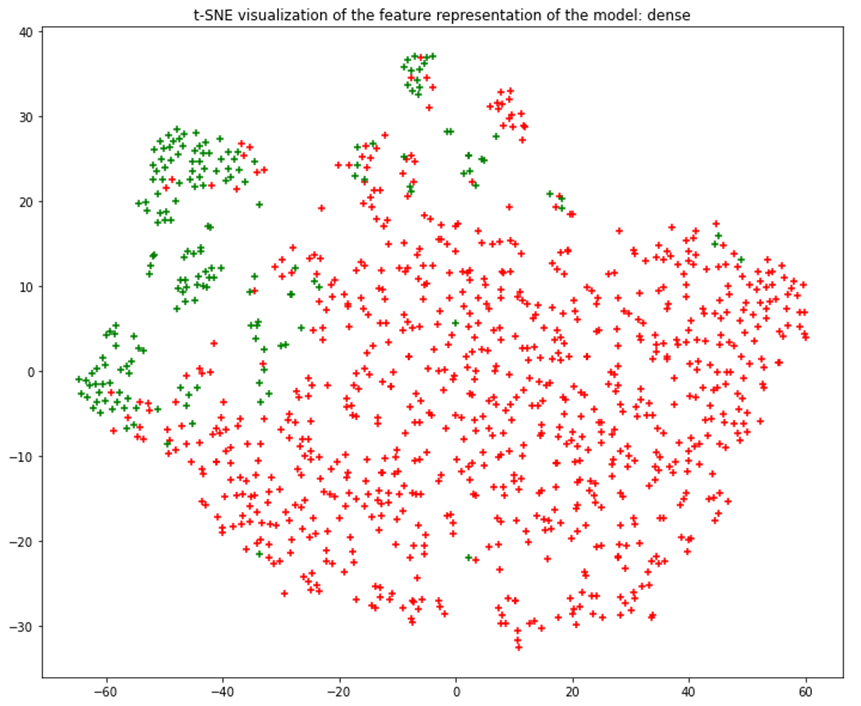}%
}%
\subcaptionbox{Second dense layer}[.40\linewidth]{%
\includegraphics[width=\linewidth]{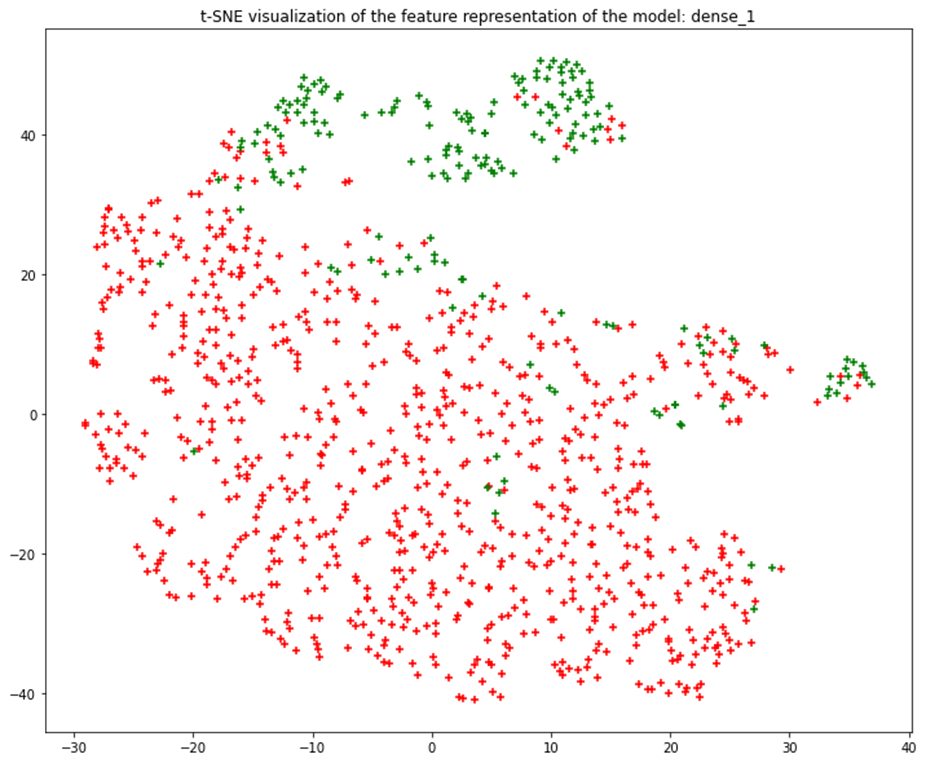}%
}
\caption{t-SNE plots for models trained on buggy data}
\label{fig:buggyDense}
\end{figure}

To assess how the quality issues in text data affect deep learning models' training behaviour, we also performed a qualitative analysis using t-SNE plots. First, we trained 30 models on clean datasets and 30 models on datasets containing concept drift. We selected concept drift for our qualitative analysis, as it affected the majority of models (see Table.~\ref{tab:rq2Table}). Then, we generated the t-SNE plots for the dense layer weights, as shown in Fig.~\ref{fig:bugfreeDense} and Fig.~\ref{fig:buggyDense}. We focus on dense layers since their decision boundaries capture complex, non-linear relationships between features and directly contribute to the final classification. Furthermore, existing literature~\cite{messaoud2022duplicate} has used the t-SNE plots of dense layers in their qualitative analysis. We discuss the outcome of our analysis below.

From Fig.~\ref{fig:bugfreeDense} and~\ref{fig:buggyDense}, we note the impact of concept drift on the dense layers of our analyzed deep learning models for duplicate bug report detection. In the first dense layer (Fig.~\ref{fig:bugfreeDense}(a)), models trained on clean data show a clear separation between the two classes (duplicate and non-duplicate), indicating effective capture of the patterns. On the other hand, models trained on buggy data (Fig.~\ref{fig:buggyDense}(a)) exhibit complex feature representations with ambiguous decision boundaries. This observation is supported by an analysis of KL Divergence, which serves as the cost function of the t-SNE plot. KL Divergence is automatically calculated and optimized during the generation of these visualizations, with lower KL Divergence values indicating a better representation of the high-dimensional data in the 2D space. For the first dense layer, the mean KL divergence for bug-free models (0.26) is lower than that of buggy models (0.32), indicating more noise in the layer representation of models affected by concept drift. The second dense layer (Fig.~\ref{fig:bugfreeDense}(b) and Fig.~\ref{fig:buggyDense}(b)) demonstrates similar trends, with bug-free models showing well-defined decision boundaries and buggy data models displaying ambiguous decision boundaries and fragmented clusters. This trend is further corroborated by the KL divergence analysis, where bug-free models exhibit a mean KL divergence of 0.14 compared to 0.24 for buggy models in the second dense layer. The increase in KL divergence for buggy models across both layers demonstrates the concept drift's impact on the network. It also highlights how concept drift progressively affects a model's ability to learn meaningful representations and make accurate predictions, ultimately impairing its performance in duplicate bug report detection.
\begin{rqbox}
\textbf{Summary of RQ2:} Quality issues in text-based data introduce many challenges for deep learning models, including abnormal weight distribution, gradient skewness, and overfitting. Moreover, missing preprocessing techniques can lead to extreme weights and skewed bias distributions, worsening the model's ability to learn effectively from the bug reports.
\end{rqbox}

\subsection{RQ3: How do data quality and preprocessing issues in metric-based data affect the training behaviour of deep learning models?}

\begin{table}
\centering
\setlength{\tabcolsep}{4pt}
\caption{Manifestations of Data Bugs in Metric-Based Models}
\label{tab:rq3Table}
\resizebox{\columnwidth}{!}{%
\begin{tabular}{|l|l|l|l|}
\hline
\textbf{Data Quality Issues} & \textbf{Sparse Parameter Updates} & \textbf{Vanishing Gradients} & \textbf{Higher Loss} \\ \hline
\multicolumn{4}{|c|}{\textbf{DeepJIT}} \\ \hline
Label Noise    & 33.33\% & 36.67\% & 23.33\% \\ \hline
Class Imbalance & 76.67\% & 83.33\% & 66.67\% \\ \hline
Concept Drift   & 13.33\% & 16.67\% & 10.00\% \\ \hline
\multicolumn{4}{|c|}{\textbf{CodeBERTJIT}} \\ \hline
Label Noise    & 53.33\% & 56.67\% & 40.00\% \\ \hline
Class Imbalance & 83.33\% & 90.00\% & 73.33\% \\ \hline
Concept Drift   & 23.33\% & 26.67\% & 16.67\% \\ \hline
\end{tabular}
}
\end{table}

\begin{table}
\centering
\setlength{\tabcolsep}{4pt}
\caption{Impact of Data Bug Symptoms on Model's Performance for Metric-Based Data}
\label{tab:metricImpact}
\resizebox{\columnwidth}{!}{%
\begin{tabular}{|l|l|l|l|l|}
\hline
\textbf{Symptoms} & \textbf{Accuracy} & \textbf{Precision} & \textbf{Recall} & \textbf{F1-Score} \\ \hline
\multicolumn{5}{|c|}{\textbf{DeepJIT}} \\ \hline
Baseline Performance & 73.14\% & 19.36\% & 71.02\% & 31.26\% \\ \hline
Sparse Parameter Updates & 69.23\% & 18.14\% & 67.86\% & 28.37\% \\ \hline
Vanishing Gradients & 64.12\% & 16.28\% & 60.34\% & 25.46\% \\ \hline
Higher Loss & 65.87\% & 17.23\% & 61.92\% & 26.18\% \\ \hline
\multicolumn{5}{|c|}{\textbf{CodeBERTJIT}} \\ \hline
Baseline Performance & 89.59\% & 36.78\% & 56.71\% & 39.45\% \\ \hline
Sparse Parameter Updates & 84.87\% & 34.23\% & 52.16\% & 35.28\% \\\hline
Vanishing Gradients & 80.24\% & 29.15\% & 46.32\% & 35.47\% \\ \hline
Higher Loss & 81.93\% & 31.26\% & 48.75\% & 37.84\% \\ \hline
\end{tabular}
}
\end{table}

\subsubsection{Impact of Data Quality}
\looseness=-1
\textbf{(a) Sparse Parameter Updates}: When we trained our models using buggy metric data, we observed sparse parameter updates in several layers. Normal learning is characterized by active neurons with regular gradient updates, whereas sparse updates to model parameters (e.g., gradients, weights, biases) suggest learning difficulties~\cite{li2024neurrev}. As shown in Table~\ref{tab:rq3Table}, 80.00\% of the models trained on data with class imbalance demonstrated sparse parameter updates. Our manual analysis shows that 50-55\% of their layers were affected, especially the fully connected dense layers, where a significant number of neurons remained inactive due to data imbalance. In contrast, models trained on balanced data enjoyed regular updates to their neurons and demonstrated consistent learning behavior. We also observed performance deterioration in the models with sparse updates. As shown in Table 3, DeepJIT witnessed 4.45-9.24\% performance drop across various metrics. CodeBERTJIT exhibited similar behaviour with a decrease of 5.27-10.57\% across various performance metrics.

\textbf{(b) Vanishing Gradients}: We observed severe gradient vanishing problem in our models trained on imbalanced datasets. According to existing studies, smaller gradients hinder information flow within a model and introduce learning difficulties~\cite{pascanu2013difficulty, goodfellow2016deep}. Vanishing gradients occur when the gradient signal becomes progressively smaller as it propagates backward through the network, preventing effective weight updates in earlier layers. This phenomenon limits the network's ability to learn long-range dependencies and complex patterns in the data. As shown in Table~\ref{tab:rq3Table}, 86.67\% of models trained on imbalanced data demonstrated vanishing gradients. Our analysis of these models revealed that gradients decreased exponentially across the layers, with input layers receiving gradients $10^6$ times smaller than output layers. This reduction significantly differs from that observed in models trained on clean data (e.g., factor of $10^2$). These vanishing gradients had a significant impact on the model's performance. As shown in Table~\ref{tab:metricImpact}, DeepJIT witnessed a performance reduction of up to 12.33-18.55\% across the various performance metrics. CodeBERTJIT exhibited similar degradation with 10.09-20.74\% decreases across all performance metrics.

\textbf{(c) Higher Loss}: When we trained the models using buggy metric data, we also observed significantly higher training losses. From the perspective of neural networks, optimal learning suggests decreasing loss values as the models progressively fit the training data~\cite{goodfellow2016deep}. Higher persistent loss indicates a model's struggle to find optimal parameters that minimize the objective function. As shown in Table~\ref{tab:rq3Table}, 70.00\% of models trained with class imbalance demonstrated this issue. These models showed stagnant losses after the first epoch, with average losses $\approx$1.50 times higher than baseline models. For instance, training loss increased from $\approx$75 to $\approx$115 in OpenStack and from $\approx$110 to $\approx$170 in QT projects. This resulted in performance deterioration, DeepJIT witnessed 9.94-16.25\% relative drop across performance metrics. CodeBERTJIT exhibited similar degradation with 4.08-15.01\% relative decreases across all performance metrics (Table~\ref{tab:metricImpact}).

\subsubsection{Impact of Missing Preprocessing Operations}
\looseness=-1
\textbf{(a) Exploding Gradients}: When we trained our models without normalizing their features, we observed extreme gradient values across multiple layers. As shown in Table~\ref{tab:rq3NormScaleTable}, 70.00\% of DeepJIT and 76.67\% of CodeBERTJIT models trained without feature normalization demonstrated exploding gradients. Our manual analysis shows that 70-80\% of their layers were affected, especially output layers, which exhibited extreme gradient values (e.g., -4.23 to 2.76), significantly deviating from the stable values (e.g., -0.95 to 0.86) observed in baseline models. We also observed notable performance deterioration in models with exploding gradients. As shown in Table~\ref{tab:rq3preprocessMetricImpact}, DeepJIT witnessed a 9.87\%-15.13\% decline across various performance metrics. CodeBERTJIT exhibited similar behavior, with metrics decreasing by 5.55\%-17.81\%.

\textbf{(b) High Variance in Weight Distribution}: We observed significant variance in the weight distributions of our models trained without normalized features. According to existing studies, balanced parameter distributions typically show moderate variance ($\sigma^2 \approx 0.89$)~\cite{lecun2002efficient}. As shown in Table~\ref{tab:rq3NormScaleTable}, 63.33\% of DeepJIT and 70.00\% of CodeBERTJIT models trained without feature normalization demonstrated high weight variance. Our analysis of these models revealed that 85-90\% of their layers were affected, particularly input and early hidden layers. These layers had high variance values (e.g., $\sigma^2 \approx 5.76$), significantly different from that of the models trained with normalized features. These high-variance weight distributions also had a noticeable impact on the model's performance, causing a 7.81\%-11.58\% and 0.43\%-11.77\% drops in various performance metrics for our baseline techniques (DeepJIT, CodeBERTJIT).

\begin{table}[h]
\centering
\caption{Impact of Missing Preprocessing in Metric-Based Models}
\label{tab:rq3NormScaleTable}
\resizebox{\columnwidth}{!}{%
\begin{tabular}{|l|l|l|}
\hline
\textbf{Missing Operation} & \textbf{Exploding Gradients} & \textbf{Weight Variance} \\ \hline
\multicolumn{3}{|c|}{\textbf{DeepJIT}} \\ \hline
Feature Normalization & 70.00\% & 63.33\% \\ \hline
Feature Scaling & 66.67\% & 60.00\% \\ \hline
\multicolumn{3}{|c|}{\textbf{CodeBERTJIT}} \\ \hline
Feature Normalization & 76.67\% & 70.00\% \\ \hline
Feature Scaling & 73.33\% & 66.67\% \\ \hline
\end{tabular}%
}
\end{table}

\begin{table}
\centering
\caption{Impact of Missing Preprocessing on Model's Performance for Metric-Based Data}
\label{tab:rq3preprocessMetricImpact}
\resizebox{\columnwidth}{!}{%
\begin{tabular}{|l|l|l|l|l|}
\hline
\textbf{Symptoms} & \textbf{Accuracy} & \textbf{Precision} & \textbf{Recall} & \textbf{F1-Score} \\ \hline
\multicolumn{5}{|c|}{\textbf{DeepJIT}} \\ \hline
Performance on Clean Data & 73.14\% & 19.36\% & 71.02\% & 31.26\% \\ \hline
Exploding Gradients & 65.92\% & 16.87\% & 63.14\% & 26.53\% \\ \hline
High Variance in Weight Distribution & 67.43\% & 17.56\% & 65.23\% & 27.64\% \\ \hline
\multicolumn{5}{|c|}{\textbf{CodeBERTJIT}} \\ \hline
Performance on Clean Data & 89.59\% & 36.78\% & 56.71\% & 39.45\% \\ \hline
Exploding Gradients & 82.14\% & 30.23\% & 48.92\% & 37.26\% \\ \hline
High Variance in Weight Distribution & 84.37\% & 32.45\% & 50.16\% & 39.28\% \\ \hline
\end{tabular}
}
\end{table}

\subsubsection{Qualitative Analysis: Impact of Class Imbalance in the Training Data on Feature Importance}

GradCAM visualization has been used in software engineering research to analyze the feature importance in defect prediction models~\cite{selvaraju2017grad, issta2021deepjit}. In our study, we also analyze GradCAM outputs to understand the impact of data quality issues on the model's training. Among four data quality issues, we found that class imbalance had the strongest impact on models trained with metric-based data (see Table~\ref{tab:rq3Table}). Thus, we conduct a qualitative analysis to better understand how class imbalance affects a model during training. Towards this, we trained 60 models in total. We divided these models into four groups of 15 models each: (a) models trained on a clean Openstack dataset, (b) models trained on a clean QT dataset, (c) models trained on an Openstack dataset with class imbalance, and (d) models trained on a QT dataset with class imbalance. We discuss our findings below.

\textbf{Buggy Models:} Our analysis of the GradCAM outputs targeting the buggy models reveals a disproportionate focus on obscure or overly specific tokens by the models. In the models trained on OpenStack dataset, tokens like `audit\_location\_generator' (0.9987) and `i62ce43a330d7ae94eda4' (0.9962) have probabilities very close to 1, while more general programming concepts like `if' (0.0023) and `return' (0.0041) have probabilities near 0. These probabilities indicate the influence of the tokens on the final output. So, the high probabilities of relevant tokens indicate that the model uses the tokens relevant to the defects in making the final decision. Similarly, the models trained on the Qt dataset exhibit high probabilities for tokens such as `Q\_OBJECT\_FAKE' (0.9976) and `qAsConst' (0.9953), while potentially salient tokens like `nullptr' (0.0037) or `delete' (0.0052) have very low probabilities. These patterns suggests that the dominant presence of negative instances (a.k.a., severe class imbalance) has likely led to overfitting, causing the models to associate rare, noisy tokens, or codebase-specific terms with software defects in the training data. Consequently, these buggy models may fail to generalize and capture the broader semantics and context of software defects across different codebases.
\begin{figure}[htbp]
    \centering
    \begin{subfigure}{\linewidth}
        \includegraphics[width=1\linewidth]{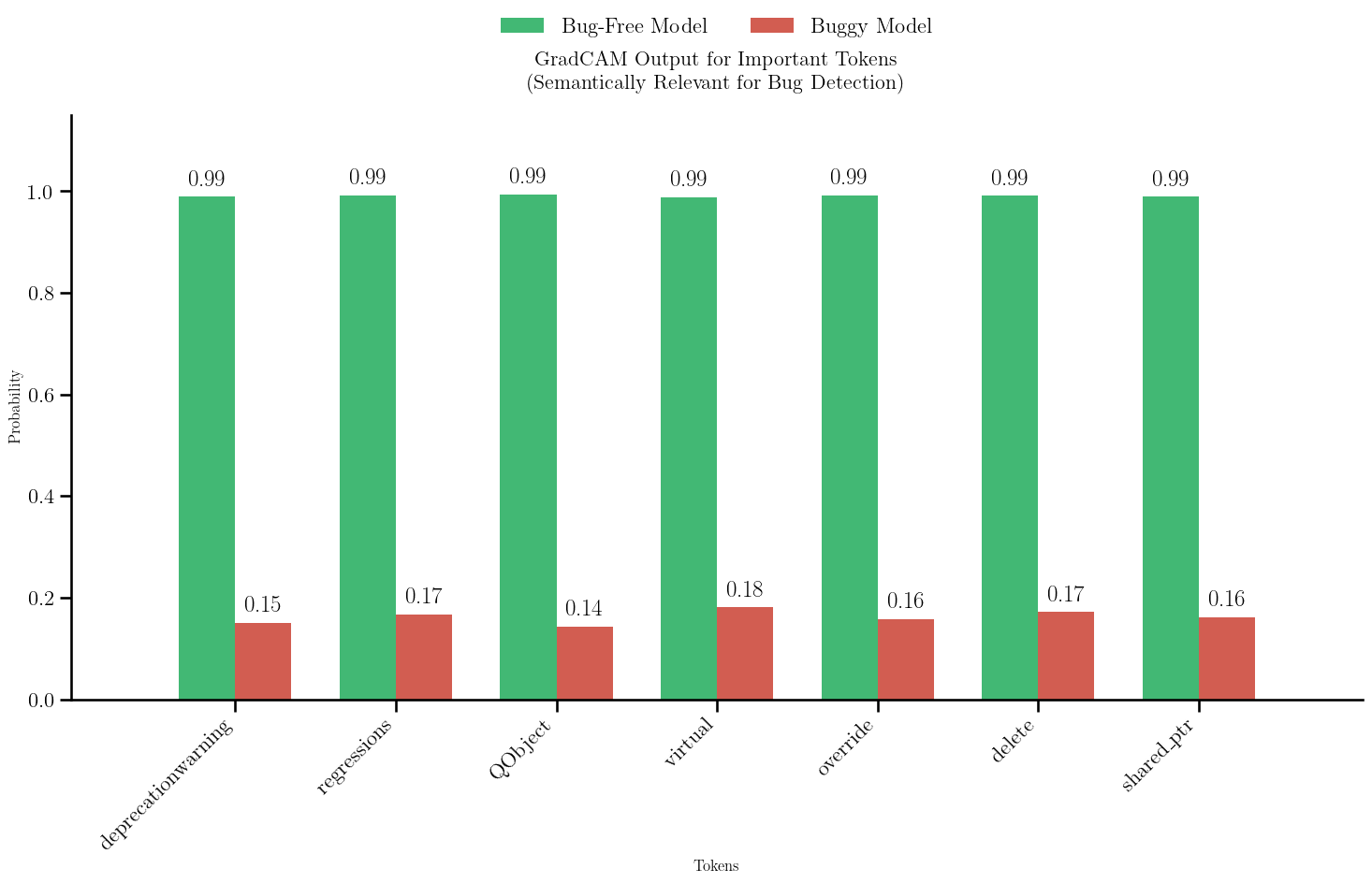}
        \caption{GradCAM outputs for high-relevance tokens demonstrating clear differentiation between bug-free and buggy models.}
        \label{fig:important-gradcam-tokens}
    \end{subfigure}
    
    \begin{subfigure}{\linewidth}
        \includegraphics[width=1\linewidth]{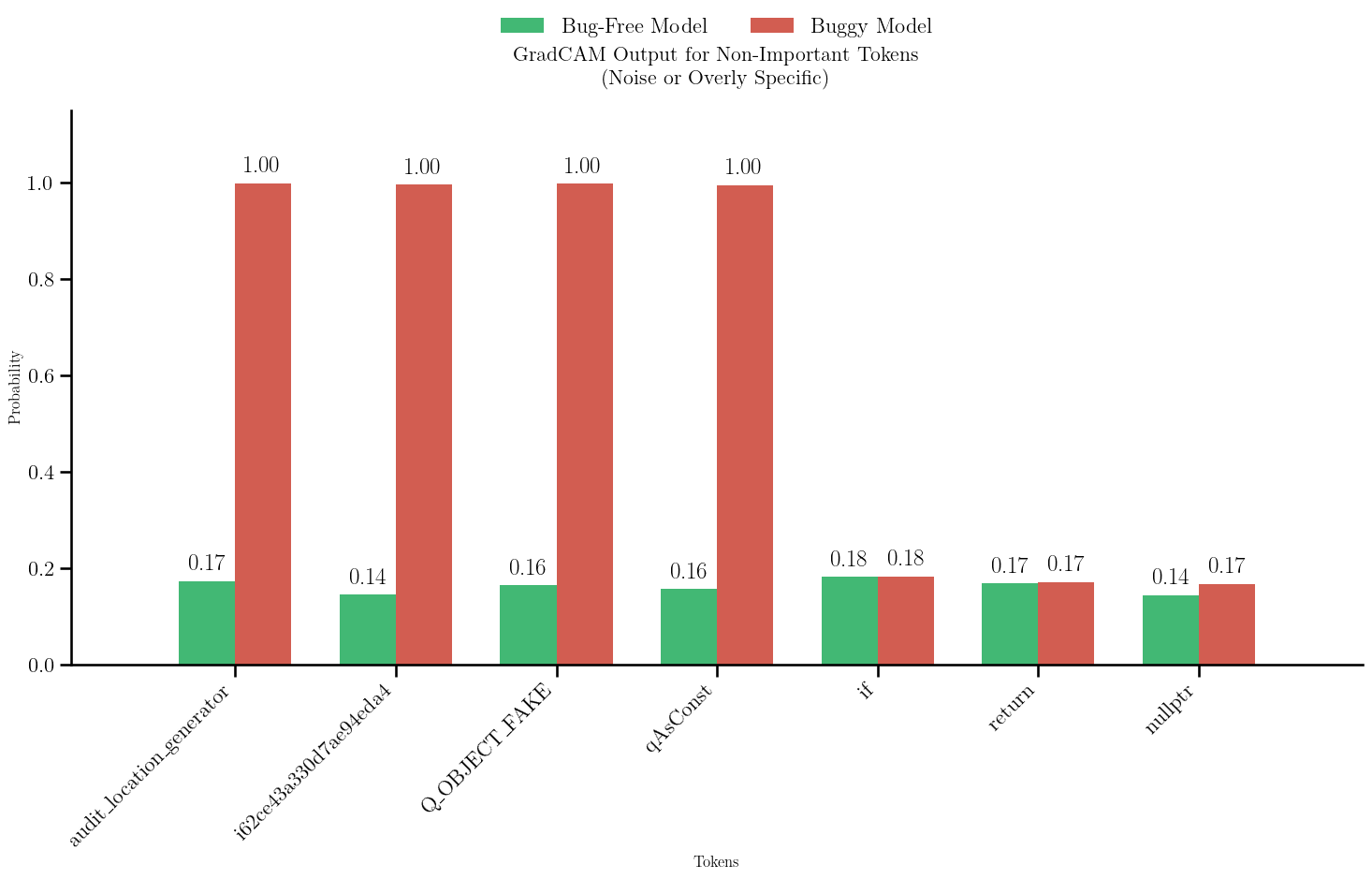}
        \caption{GradCAM outputs for low-relevance tokens showing opposite trends between bug-free and buggy models.}
        \label{fig:non-important-gradcam-tokens}
    \end{subfigure}
    
    \caption{Analysis of GradCAM activation patterns across token relevance. The bug-free model (green) shows high probabilities for semantically relevant tokens and low for irrelevant ones, while the buggy model (red) exhibits inverse behaviour, validating the model's ability to identify bug-indicative features.}
    \label{fig:gradcam-analysis}
\end{figure}

\textbf{Bug-free Models:} In contrast, according to GradCAM output, the bug-free models focus on a wider variety of relevant technical terms and phrases in both OpenStack and Qt codebases. For OpenStack, tokens such as `vendor-data' (0.9871), `xenapinfs-glance-integration' (0.9923), `live\_migr-ate' (0.9905), `deprecationwarning' (0.9889), `regressions' (0.9917), and `backward' (0.9894) have probabilities close to 1. These terms intuitively relate to areas where bugs occurred in OpenStack code, such as migrations, deprecations, regressions, and vendor integrations. Similarly, for Qt, we observe a strong focus on both framework-specific and general programming concepts. Tokens like `QObject' (0.9934), `connect' (0.9912), `emit' (0.9901), `virtual' (0.9887), and `override' (0.9923) have probabilities close to 1. Additionally, memory-related terms such as `new' (0.9876), `delete' (0.9908), and `shared\_ptr' (0.9892) also show high probabilities, reflecting the importance of memory management in C++ code. These observations suggest that the bug-free models learn to focus on both codebase-specific concepts and general programming patterns that might have historically been associated with software defects. With balanced training data, they might have developed more generalizable representations that meaningfully capture the semantics of software defects across different codebases and frameworks.

The GradCAM visualization in Figure \ref{fig:gradcam-analysis} shows a striking contrast in how bug-free and buggy models process code tokens. For semantically relevant tokens, the bug-free models demonstrate consistently high probabilities (e.g., 0.99 for 'deprecationwarning', 'regressions', 'QObject', and 'virtual'), while the buggy models show low probabilities (0.14-0.18). Conversely, for non-important tokens, the pattern inverts dramatically - the buggy models assign maximum probability (1.00) to noise tokens like 'i62ce43a330d7ae94eda4'. Whereas, the bug-free models maintain appropriately low probabilities for the same tokens (0.14-0.18). Moreover, both models show similar low probabilities (0.17-0.18) for basic programming tokens, suggesting the bug-free models have a better understanding of defective and non-defective tokens. The probabilities across the salient and noisy tokens demonstrate that the bug-free models have learned to effectively discriminate between salient features and noise, while the buggy models are skewed toward overly specific or irrelevant tokens, limiting their generalization capability across different codebases.

\begin{rqbox}
\textbf{Summary of RQ3:} The presence of data quality issues in metric-based data introduces several challenges for deep learning models, including sparse parameter updates, vanishing gradients, and higher loss. Moreover, inadequate preprocessing techniques can lead to exploding gradients and high variance in weight distribution, further hindering the models' ability to learn from the data effectively.
\end{rqbox}

\subsection{RQ4: How well do our findings on data quality and preprocessing issues generalize to other code-based, text-based, and metric-based datasets?}

To evaluate the generalizability of our findings, we conducted experiments on six additional datasets: D2A and Juliet (code-based), Spark and Mozilla (text-based), and Go and JDT (metric-based). We present our findings below.

In code-based datasets, we found that $81.67\%$ of models exhibited near-zero biases, $85.00\%$ showed smaller weights, and $68.33\%$ experienced gradient instability when trained on data with label noise or concept drift. For text-based datasets, $85.00\%$ of models showed abnormal weight distribution, $73.33\%$ demonstrated gradient skewness, and $85.00\%$ exhibited overfitting when trained on data with concept drift or class imbalance. In metric-based datasets, $80.00\%$ of models showed sparse parameter updates, $86.67\%$ experienced vanishing gradients, and $70.00\%$ demonstrated poor optimization when trained on data with label noise or class imbalance.

When models were trained without appropriate preprocessing of data, we also observed similar patterns across all dataset types. In code-based datasets, $70.00\%$ of models trained without proper preprocessing showed slow convergence, and $56.67\%$ exhibited extreme bias values. For text-based data, $68.33\%$ of models trained on unprocessed data showed extreme weight values, while $55.00\%$ exhibited skewed bias distributions. In metric-based datasets, $70.00\%$ of models trained without preprocessing demonstrated exploding gradients, and $56.67\%$ showed high variance in weight distribution. All percentages observed in these new datasets align closely with our original findings, varying within a $1$-$5\%$ margin. These consistencies in observations across diverse datasets underscore the generalizability of our findings, highlighting the persistent impact of data quality issues and the importance of proper preprocessing in model training for software engineering tasks.

\subsubsection{Model's Behaviour on Clean Datasets} 
After addressing data quality and preprocessing issues and retraining the above models, we observed significant improvements in the behaviour of the deep learning models and their performance across all types of datasets: code-based, text-based, and metric-based. The impacts of data bugs were significantly diminished, and the models exhibited stable and normal behaviours during their training.

In code-based datasets, we found only $5.00\%$ of models showing signs of near-zero biases, compared to $81.67\%$. The quality of model weights improved substantially, with only $6.67\%$ showing the symptom of smaller weights. Gradient stability also increased, with only $8.33\%$ of models demonstrating unstable gradients, down from $68.33\%$ in the presence of data quality issues. For text-based datasets, the chance of models having abnormal weight distribution decreased dramatically, with only $6.67\%$ of models showing this behavior, compared to $85.00\%$ before. Gradient skewness was largely prevented, with only $8.33\%$ of models demonstrating skewed gradient distributions. Overfitting was significantly reduced, with only $5.00\%$ of models showing this issue, down from $85.00\%$ previously. In metric-based datasets, sparse parameter updates were nearly prevented, with only $6.67\%$ of models showing any signs of this issue. The vanishing gradient problem was significantly mitigated, with only $8.33\%$ of models demonstrating this issue. Poor optimization was largely addressed, with only $11.67\%$ of models showing this problem, compared to $70.00\%$ previously.

When models were trained with proper preprocessing techniques, we observed substantial improvements across all dataset types. In code-based datasets, we found only $3.33\%$ of models showing unstable convergence patterns, compared to $70.00\%$ previously. The quality of the bias distribution improved significantly, with only $5.00\%$ showing extreme bias values. For text-based datasets, the occurrence of extreme weights decreased dramatically, with only $5.00\%$ of models showing this behaviour, compared to $68.33\%$ before. Bias distribution issues were largely eliminated, with only $6.67\%$ of models demonstrating skewed distributions. In metric-based datasets, exploding gradients were nearly eliminated, with only $5.00\%$ of models showing signs of this issue. The high variance in weight distribution was significantly mitigated, with only $6.67\%$ of models demonstrating this issue.

\subsubsection{Statistical Signifiance Tests}
To assess the generalizability of our findings, we performed statistical significance tests with the following hypotheses.
\begin{align*}
H_0&: \text{The presence of data quality issues and model} \\
   &\phantom{:} \text{symptoms are independent}\\
H_1&: \text{The presence of data quality issues and model} \\
   &\phantom{:} \text{symptoms are dependent}
\end{align*}

We selected McNemar’s test~\cite{pembury2020effective} for the statistical tests due to our experimental design and paired nominal data, as it specifically evaluates changes in binary outcomes (presence/absence of symptoms) within the same models. While McNemar's test establishes statistical significance to test our null hypothesis that data quality issues and model symptoms are independent, the odds ratio measures the strength of their dependence~\cite{szumilas2010explaining}. In our analysis, an odds ratio of 1 would support our null hypothesis, which indicates that there is no association between data quality issues and model symptoms. Conversely, odds ratios greater than one would support our alternative hypothesis, suggesting that the presence of data quality issues increases the likelihood of observed symptoms. Moreover, odds ratios below 1 would indicate that data quality issues decrease the likelihood of model symptoms.

McNemar's test results consistently rejected the null hypothesis across all symptoms for data quality and data preprocessing issues ($p < 0.01$). The odds ratios, measuring the strength of association between data quality issues and symptoms, were substantially greater than 1 in all cases, ranging from $15.53$ for abnormal weight distribution to $806.10$ for gradient skewness, strongly supporting our alternative hypothesis. The most pronounced dependencies were observed in near-zero biases, gradient instability, and overfitting ($p < 0.0001$), indicating that these symptoms were exclusively present when data quality issues existed. Even the weakest association, found in the symptom "high variance in weight distribution" ($p = 0.0098$), demonstrated a statistically significant association between the symptom and data quality issues.

These results demonstrate that addressing data quality issues and applying appropriate proper preprocessing operations to training data can significantly improve model training for various software engineering tasks. The consistent improvements observed across different dataset types further validate the generalizability of our findings and underscore the critical importance of data quality and preprocessing in machine learning applications for software engineering tasks.

\begin{rqbox}
\textbf{Summary of RQ4:} Our findings on data quality and preprocessing issues generalize consistently across diverse software engineering datasets, including D2A and Juliet (code-based), Spark and Mozilla (text-based), and Go and JDT (metric-based). The repeated emergence of similar patterns and enhancements after addressing the data quality issues across multiple datasets increases confidence in our study findings.
\end{rqbox}

\section {Discussion}

\subsection{Impact of Data Quality Issues for Generation Tasks}
Generation tasks have gained significant attention in the past few years, thanks to the emergence of large language models. As shown in existing studies~\cite{yang2022dlinse}, these tasks are the second most prevalent type of task in software engineering. To understand the impact of data bugs on a generation task, we conduct a prototype experiment that investigates how data quality and preprocessing issues affect code summarization. We utilized two well-known datasets, FunCom and TLCodeSum, which have been used in prior work~\cite{mu2023developer, gao2023makes, gros2020code} and are known to contain data quality issues~\cite{shi2022we, vitale2025optimizingdatasetscodesummarization}.

To address these issues from the datasets, we employed two distinct techniques from prior work~\cite{shi2022we, vitale2025optimizingdatasetscodesummarization}. First, we used SIDE-based filtering, which measures the coherence between code snippets and comments using contrastive learning, and filters out instances with low SIDE scores. Additionally, we employed CAT, a rule-based preprocessing tool that leverages an existing taxonomy~\cite{shi2022we} to identify the comment-related and code-related noise in the dataset. The tool then applies two primary strategies to clean the data. The first strategy is to fix minor issues in comments by removing inappropriate or noisy content, such as boilerplate text and block-level comments. The second strategy is to completely drop instances where the code itself is structurally problematic, for example, removing code-comment pairs for simple getter and setter methods. We use CodeT5+ as the base model, following the approach employed in existing studies \cite{vitale2025optimizingdatasetscodesummarization}.

We trained three variants of CodeT5+ to evaluate the effects of data quality and preprocessing. These variants are described below: 
\begin{itemize}
    \item \textbf{Clean}: The model was trained on datasets that had been carefully cleaned using both the SIDE filtering tool and the CAT preprocessing tool. 
    \item \textbf{Buggy Data}: The model was trained on the datasets without SIDE filtering. This allowed us to assess the impact of raw, unfiltered data containing inherent quality issues (e.g., outdated code-comment pairs). 
    \item \textbf{Missing Preprocessing}: The model was trained on datasets that were not preprocessed using the CAT preprocessing tool.
\end{itemize}

When we trained the models with buggy data, we observed a significant difference in their gradient behaviours. Gradient histograms of the models trained on clean data demonstrated tightly concentrated values. For example, a representative gradient distribution from the clean dataset shows values predominantly in the range of approximately $1.12 \times 10^{-5}$ to $7.44 \times 10^{-5}$. This central clustering indicates stable and consistent parameter updates during backpropagation. In contrast, models trained on buggy data exhibit significant instability in their gradients. Their gradient distributions are often wider and irregular, with a noticeable presence of outliers. For instance, a gradient histogram from the low-quality dataset shows a broader distribution (e.g., $-0.000157$ to $8.54 \times 10^{-6}$). This dispersed distribution indicates less stable gradient updates and potential issues during the optimization process. We observed this behaviour in classification-based tasks as well, suggesting the generalizability of our findings.

The absence of a preprocessing step also significantly impacts a model's training process. Our analysis of parameter distributions reveals the effect of missing preprocessing on model training. Models trained with adequate preprocessing demonstrate well-defined and stable weight distributions. A representative parameter histogram from our clean model shows that values are evenly distributed within a contained range (e.g., from 0.098 to 0.383). This indicates that the model has learned a structured and balanced set of weights, which is essential for effective generalization. Conversely, a model trained without adequate preprocessing shows a broader and more dispersed weight distribution, with values spanning a wider range (e.g., from 0.061 to 0.766). This pattern, also seen in classification tasks, further supports the generalizability of these findings.

\subsection{Implications for Stakeholders}
The findings of our study have significant implications for various stakeholders involved in developing and deploying deep learning models targeting software engineering tasks. We discuss these implications in detail below.

\textbf{Researchers:} Researchers can reuse our dataset and leverage our empirical insights concerning label noise, concept drift, and class imbalance to develop robust data cleaning and preprocessing techniques. Our findings on the impact of missing preprocessing operations can also guide researchers in designing appropriate preprocessing pipelines tailored to different data types (code-based, text-based, and metric-based). Additionally, our analysis of how data quality issues affect the training of DL models can inform the development of more effective debugging techniques for DL models. Furthermore, researchers can integrate the explainable AI (XAI) techniques demonstrated in our study, such as attention visualization, t-SNE, and GradCAM, into their methodologies to better understand model behaviour and the impact of data quality issues. Finally, the researchers can use the symptoms of data bugs as verification criteria to enhance the reproducibility of deep learning bugs, as highlighted by existing literature~\cite{shah2025towards}.

\textbf{Software Engineers:} Software engineers can leverage our insights and implement continuous data monitoring systems that track properties such as weight distributions, gradient patterns, and learning stability across different layers of a model. These systems can be used to automatically detect anomalies in the training process and data distributions, enabling early identification of potential data quality issues before the models are deployed in real-world settings. Our insights on the effects of concept drift can be used to establish processes for regular model retraining and deployment, ensuring their relevance over time. Moreover, the XAI techniques from our study can help software engineers debug and analyze model behaviour, identify potential biases, and address any issues before their model deployment.

\textbf{Deep Learning Engineers:} Deep learning engineers can leverage our empirical findings to establish automated data quality assurance and design validation mechanisms and integrate them into MLOps pipelines. The identified data quality issues and their manifestations can serve as the basis for designing effective validation rules, statistical tests, and domain-specific constraints. Furthermore, our study emphasizes the importance of robust preprocessing pipelines, which can guide deep learning engineers in building modular and extensible preprocessing components that adapt to different data types and evolving data characteristics. Our findings enable deep learning engineers to develop optimized training procedures with built-in validation mechanisms to ensure stable model convergence. Additionally, they can implement comprehensive monitoring and alerting systems that track critical indicators like weight distributions, gradient flows, and learning patterns, allowing rapid detection and response to potential data quality issues during model training.

\subsection{Building Robust Data Pipelines for Deep Learning Model Training}
Based on our findings, we recommend a systematic data pipeline for training models targeting software engineering tasks. It has four steps: data collection and versioning, quality validation, preprocessing, and monitoring throughout the training process.

First, we recommend establishing data collection and versioning protocols to promote reproducibility and traceability in the process. In this step, data should be collected from appropriate sources: code-based data from GitHub repositories and CI/CD logs, text-based data from bug trackers and commit messages, and metric-based data from static analysis tools. Organizations should implement provenance tracking through comprehensive metadata logging and data version control integration for systematic versioning. To maintain data integrity, raw data should be preserved in immutable formats such as Parquet files.

Second, we recommend implementing quality validation mechanisms for the captured data above. This phase can incorporate three key validation steps: label noise detection using confidence learning~\cite{northcutt2021confident}, concept drift monitoring through time-series analysis and drift detection algorithms, and class imbalance assessment using distribution analysis and adaptive sampling strategies.

Third, developing type-specific data preprocessing strategies for different data modalities is recommended. For text-based data, data should be extracted in chronological order from SE documentation, segmented into appropriate units, noisy content should be removed, and standard NLP preprocessing with tokenization should be applied. For code-based data, ASTs should be generated, code should be normalized and standardized, and token-based representations should be created. For metric-based data, duplicates should be eliminated, missing values should be handled, and feature normalization and scaling should be applied to ensure consistent ranges.

Finally, we recommend establishing Training Process Monitoring to implement comprehensive quality control during model training. The pipeline should continuously analyze parameter distributions through IQR and Z-score analyses for outlier detection or skewness and kurtosis measurements. Layer-wise distribution monitoring should track the behaviors of individual network components and compare them against the established thresholds. For weights and biases, both extremely small magnitudes (indicating potential vanishing gradients) and unusually large values (suggesting exploding gradients) should trigger monitoring alerts based on predefined thresholds. Based on observed anomalies, the model training can be interrupted and relevant details can be reported. The idea is to help subsequent analysis including root cause analysis, data repair, and re-training the model.

Based on our analysis across multiple datasets and model architectures, such a pipeline has the potential to help practitioners detect and address potential issues throughout the training process, ultimately enhancing model performance and robustness.

\subsection{DataGuardian: Prototype for identifying data quality issues}
To validate our findings and demonstrate their practical utility, we developed Data-Guardian~\cite{replicationPackage}, a monitoring system that can incorporate our parameter thresholds in a DL model for detecting data quality issues during model training. Our system continuously tracks parameter distributions, gradients, and layer-wise behaviours through a comprehensive set of statistical measures, including IQR, skewness, kurtosis, and distribution stability metrics.

DataGuardian offers flexibility in its monitoring approach through configurable thresholds. Users can either utilize our pre-defined thresholds, which were empirically determined through extensive experimentation, or define custom thresholds based on their specific requirements and domain knowledge. At the parameter level, it tracks weights and biases for symptoms such as abnormal weight distribution, overfitting, and skewed bias distributions. At the gradient level, Data-Guardian analyzes the flow of gradients through the network to detect problems such as gradient skewness, gradient instability, and vanishing or exploding gradients, which often indicate underlying data quality problems.

We evaluated DataGuardian's effectiveness using multiple datasets across different domains: Juliet for code-based data, Mozilla for text-based data and JDT for metric-based data, using CodeBERT-based baseline models. They were not part of our threshold-related analysis. In experiments with buggy data, DataGuardian successfully identified training problems and their root causes using the our established thresholds in 12/15 runs for code-based data (Juliet), 11/15 for text-based data (Mozilla), and 12/15 for metric-based data (JDT). For scenarios with missing preprocessing steps, the system accurately detected training issues and identified the lack of preprocessing as the root cause in 10/15, 11/15, and 11/15 runs, respectively, for each dataset.

Our prototype also implements customizable interruption mechanism that can interrupt model training when certain pre-defined conditions are met. Users can also choose their thresholds concerning layer coverage of the encountered issues. This flexible approach helps distinguish between isolated anomalies (e.g., vanishing gradients) and systematic data quality problems while allowing users to balance between early detection and false positives. Upon detecting issues, Data-Guardian generates detailed diagnostic reports with specific metrics, potential root causes, and suggested mitigation strategies based on existing literature.

\section{Related Work}
\looseness=-1
Data quality is crucial for the performance and reliability of deep learning models in software engineering. Previous studies have investigated the effects of data quality issues like label noise~\cite{frenay2013classification, frenay2014comprehensive, nie2023understanding, croft2023data}, class imbalance~\cite{li2022robust, liu2022comparative}, data duplication~\cite{zhang2023duplicate, croft2023data, lopez2024inter}, and concept drift~\cite{zhang2023duplicate, lai2021towards, schelter2015challenges} on deep learning models.

Recent studies have also analyzed the impact of data quality issues on deep learning models for software engineering tasks. Wu et al.~\cite{wu2021data} investigated mislabeled instances in five publicly available datasets commonly used for security bug report prediction, including Chromium, Ambari, Camel, Derby, and Wicket, where they identified 749 security bug reports incorrectly labelled as non-security. They also reported significant performance improvements for a retrained model that was trained on correctly labelled data. Tantithamthavorn et al.~\cite{tantithamthavorn2015impact} analyzed over 3,900 issue reports from Apache projects and demonstrated that label noise can significantly impact the recall of a model. They also reported a significant improvement in the recall ($\approx$60\%) for a retrained model on cleaned datasets. Kim et al.~\cite{kim2011dealing} evaluated the impact of intentionally injected noise into the datasets for defect prediction models through controlled experiments. They found that when $\approx$30\% of the data was mislabeled, the performance of defect prediction models decreased significantly, which highlights the models' sensitivity to label noise. Fan et al.~\cite{fanimpact} investigated mislabeled changes in just-in-time defect prediction datasets and found that certain labelling approaches can lead to performance reduction of up to 5\%. Xu et al.~\cite{xu2023data} leveraged adversarial learning to improve the data quality of obsolete comment datasets, which led to an improvement in the accuracy of multiple existing models by 18.1\%. Croft et al.~\cite{croft2023data} conducted a systematic analysis of data quality in software vulnerability datasets and found out that 71\% of the labels are incorrect and 17-99\% of data points are duplicated across four state-of-the-art datasets. Furthermore, they also analyzed the impact of the data quality issues on vulnerability detection. They found that the model's performance dropped by 65\% when trained on clean data, which shows how the duplication and mislabelling of the data points can lead to inflated results.

In summary, the existing studies have examined the impact of data quality on model performance. However, a clear understanding of how these issues affect the training behaviours of a DL model remains limited. Most existing studies focus on specific data types or individual quality issues, preventing a comprehensive understanding of how data quality impacts deep learning models. Besides, how missing preprocessing operations can affect model behaviour is not well understood to date. Our study addresses these gaps in the literature through a systematic analysis of 900 models and 12 datasets targeting software engineering tasks.

Our work investigates how data quality and preprocessing issues affect the training behaviours of deep learning models. We adopt a comprehensive approach by examining three major issues - label noise, class imbalance, and concept drift - across software engineering data in code, text, and metric formats. We also demonstrate the generalizability of our findings through the reproduction of our findings on separate datasets, which were not used in our primary analysis. This multifaceted approach informs us of the impact of data bugs on deep learning models targeting software engineering tasks, addresses the gaps in the existing literature and encourages future efforts for appropriate debugging solutions.

\section{Threats to Validity}

One potential threat to the \textit{internal} validity of our findings is the choice of baseline models and datasets. Although we followed systematic criteria for selection (Section 3.1, Section 3.2), there may be other relevant models or datasets not considered. We mitigate this threat by carefully documenting our selection criteria and validating our findings across diverse datasets within each data type category (code-based, text-based, and metric-based). Another threat is the potential for errors or biases in our analysis techniques. We mitigate this threat by using analysis techniques which are well-established in the literature and triangulating our findings through qualitative analysis (using XAI techniques) and quantitative analysis (using gradients, weights, and biases). Another threat is the subjective nature of our qualitative analysis, particularly in interpreting attention weights for code comprehension, t-SNE visualizations, and GradCAM results. We mitigated this by following established guidelines for analysis from previous studies~\cite{issta2021deepjit, fu2022linevul, messaoud2022duplicate}.

The main threat to \textit{external} validity is the generalizability of our findings to software engineering tasks or data types that were not those considered. Our analysis is based on three tasks -- vulnerability prediction, defect prediction and duplicate bug report detection. The impact of data quality and preprocessing issues may manifest differently in other tasks or application domains. Additionally, we focused on specific types of data quality issues, which may limit the generalizability of our findings. To mitigate these threats, we selected prevalent tasks, data types, and data quality issues based on comprehensive surveys and prior studies~\cite{yang2022dlinse, zhang2023duplicate, croft2023data, li2022robust, liu2022comparative} and used established benchmark datasets from different projects and domains.

The primary threat to \textit{conclusion} validity comes from the inherent stochasticity in deep learning model behaviours. To account for this, we run each model multiple times and collectively analyze our findings in an aggregate manner across different data types and model behaviours, examining patterns in training dynamics (through gradients, weights, and biases). Our conclusions are drawn from these consistent, aggregate patterns observed in model behaviours when encountering data quality issues.

The primary threat to \textit{conclusion} validity is the potential for statistical errors or violations of assumptions in our quantitative analysis. While we followed best practices and ran multiple trials to account for stochasticity, there may be inherent limitations or biases that could affect the accuracy or generalizability of our conclusions. Additionally, our conclusions are based on specific experiments and analysis techniques, and other approaches or methodologies could yield different or complementary insights. Regarding \textit{reproducibility}, we acknowledge several potential threats. Environmental factors such as hardware specifications, software versions, and operating systems may influence results. To mitigate these threats, we provide detailed documentation of our experimental setup and a comprehensive replication package containing all source code, datasets, and configuration files used in our study~\cite{replicationPackage}.

\section {Conclusion and Future Work}

Deep learning systems suffer from data bugs that can significantly impact their reliability and trustworthiness. These bugs are particularly challenging to detect and resolve as they originate from various sources, including training data quality issues and preprocessing errors. In this study, we investigate the impact of data quality and preprocessing issues on the training behaviours of deep learning models targeting software engineering tasks. We target three prevalent data types: code-based, text-based, and metric-based data in the context of three tasks: vulnerability detection, duplicate bug report detection, and defect prediction. First, we replicated three state-of-the-art baselines, introducing data quality issues, omitting preprocessing operations and capturing their training-time behaviours using advanced tools (e.g., Weights \& Biases). We then employed advanced explainable AI techniques to gain insights into the models' learning processes and internal representations. Our key findings shed light on the impacts of data quality issues, such as label noise, concept drift, and class imbalance. These issues can lead to near-zero biases, smaller weights, gradient instability during model training, slower convergence, overfitting to noisy patterns, vanishing gradients, higher loss, and impaired generalization capabilities. Future studies can reuse the same methodology and derive the impact of data quality and preprocessing issues on generation-related tasks such as code generation, code completion and code representation.

Additionally, the absence of necessary preprocessing operations can result in errors in extreme weights, skewed bias, and exploding gradients. This study emphasizes the critical importance of high-quality data and robust preprocessing pipelines for effective training of deep learning models in software engineering tasks. Our derived insights serve as a foundation for future work in developing data cleaning techniques, continuous data quality monitoring systems, and automated preprocessing pipelines tailored to different data types within the field of software engineering. Our findings can benefit debugging solutions by highlighting potential data-related issues that may impact model performance, developing more robust preprocessing techniques, and informing the design of automated quality assessment tools for machine learning pipelines in software engineering. These insights can also help practitioners identify and mitigate common data quality issues before negatively impacting model training and deployment.

\section*{Declarations}
\noindent \textbf{Funding and/or Conflicts of interests/Competing interests}\\
This work was supported by the Natural Sciences and Engineering Research Council of Canada (Discovery Grant RGPIN‑03236). The authors also declare they have no conflict of interest.\\
\\
\noindent \textbf{Data Availability Statement}\\
All the data generated or analyzed during this study are available on Zenodo to help reproduce our results~\cite{replicationPackage}.\\
\\
\noindent \textbf{Ethical Approval}\\
Not applicable.\\
\noindent
\\
\noindent \textbf{Informed Consent}\\
Not applicable.\\
\noindent
\\
\noindent \textbf{Author Contributions}
\begin{itemize}
    \item \textbf{Mehil B. Shah:} Conceptualization, Methodology, Implementation, Investigation, Formal analysis, Visualization, Writing - original draft
    
    \item \textbf{Mohammad Masudur Rahman:} Conceptualization, Supervision, Writing - review \& editing, Validation, Conceptual guidance
    
    \item \textbf{Foutse Khomh:} Supervision, Writing - review \& editing, Validation, Conceptual guidance
\end{itemize}
\begingroup
\small
\bibliographystyle{plainurl}
\bibliography{bibliography}
\endgroup

\newpage
\appendix
\section*{Appendix A: Sample Output for Data Guardian}
\begin{figure}[h]
    \centering
    \includegraphics[width=1\linewidth]{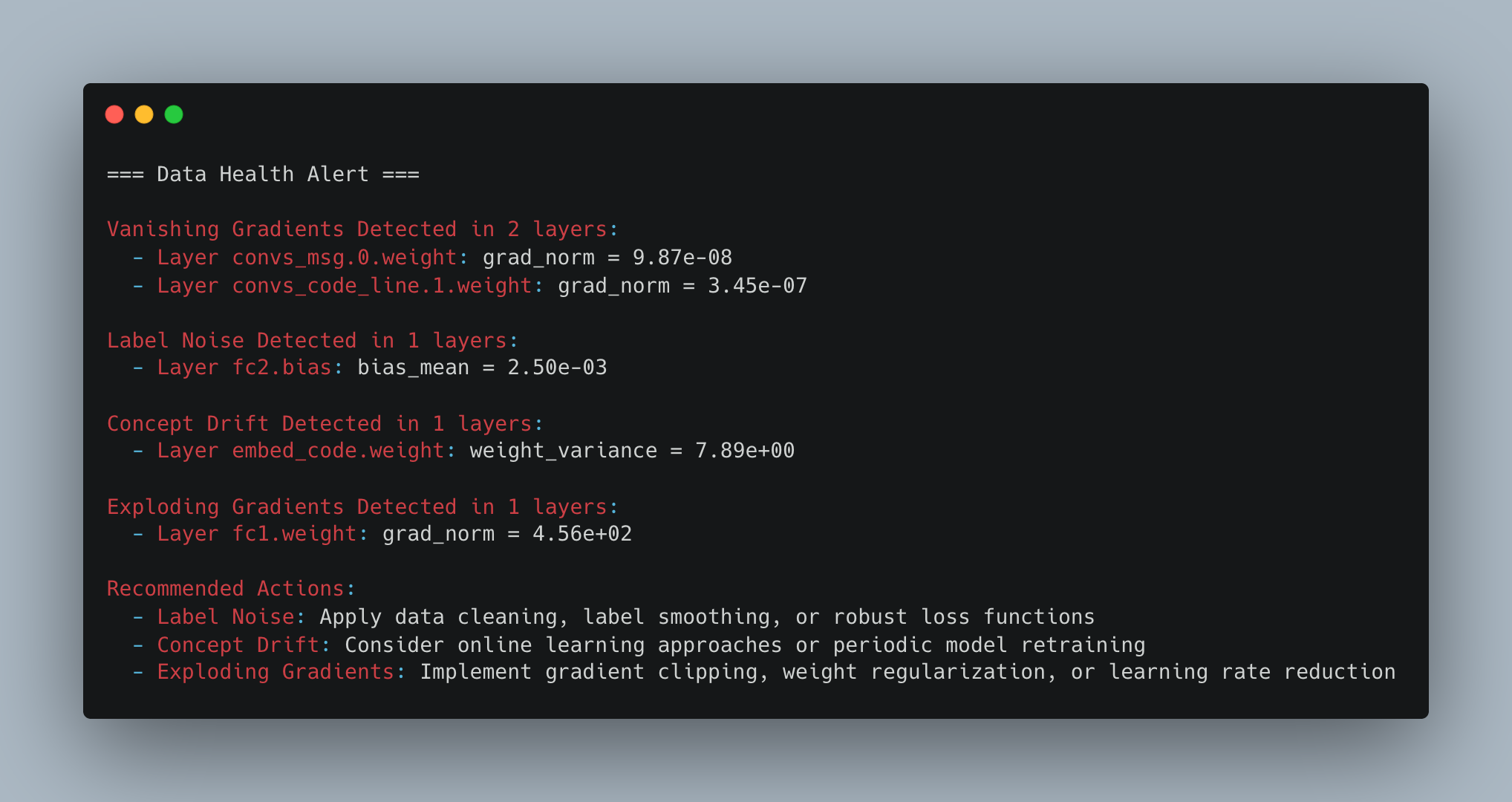}
    \caption{Sample output from Data-Guardian}
    \label{fig:data-guardian}
\end{figure}

The Data Guardian system provides real-time monitoring of neural network training processes by detecting common training pathologies. As shown in Figure \ref{fig:data-guardian}, the system identifies several critical issues that can significantly impact model performance:

\textbf{Vanishing Gradients:} This occurs when gradient values become extremely small during backpropagation, preventing effective weight updates in earlier network layers. The example shows this issue in two convolutional layers with gradient norms of 9.87e-08 and 3.45e-07, respectively. Without intervention, these layers would essentially stop learning.

\textbf{Label Noise:} Detected in the fc2.bias layer with a bias mean of 2.50e-03, this indicates potential inconsistencies in training labels. Label noise can lead to decreased generalization performance and longer convergence times as the model attempts to fit to contradictory patterns in the data.

\textbf{Concept Drift:} The embed\_code.weight layer exhibits high weight variance (7.89e+00), suggesting that the statistical properties of the input data or target variables may be changing over time. This phenomenon often occurs in production systems where input distributions evolve beyond what was present in the original training data.

\textbf{Exploding Gradients:} At the opposite extreme of vanishing gradients, this issue manifests as excessively large gradient values—in this case, a gradient norm of 4.56e+02 in the fc1.weight layer. Exploding gradients can cause numerical instability, weight updates that overshoot optimal values, and in severe cases, NaN values that halt training entirely.

For each detected issue, Data Guardian automatically suggests appropriate remediation strategies, ranging from architectural adjustments to regularization techniques. These automated alerts allow machine learning engineers to intervene promptly during training, potentially saving significant computational resources and engineering time that would otherwise be spent debugging failed training runs.

\end{document}